\documentclass[11pt]{article}

\usepackage[colorlinks,citecolor=blue,bookmarks=true]{hyperref}
\usepackage[dvipsnames]{xcolor}
\usepackage{tikz}
\usetikzlibrary{arrows.meta,decorations.pathreplacing}

\usepackage{amsmath} % AMS Math Package
\usepackage{amsthm} % Theorem Formatting
\usepackage{amssymb}	% Math symbols such as \mathbb
\usepackage{graphicx} % Allows for eps images
\usepackage{multirow}
\usepackage[dvipsnames]{xcolor}
\usepackage[dvips,letterpaper,margin=1in]{geometry}
\usepackage[capitalize,noabbrev]{cleveref}
\usepackage{subcaption}

\usepackage[utf8]{inputenc}
\usepackage[english]{babel}

\usepackage{algorithm}
\usepackage{algpseudocode}
\makeatletter
\renewcommand{\ALG@name}{Algorithm}
\makeatother

\usepackage{bm}
\usepackage{bbm}
\usepackage{diagbox}
\usepackage{mathtools}
\usepackage{ytableau}

\usepackage{tabularx, booktabs}

\newtheorem{theorem}{Theorem}[section]

\newtheorem{lemma}[theorem]{Lemma}
\newtheorem{corollary}[theorem]{Corollary}
\newtheorem{proposition}[theorem]{Proposition}
\newtheorem{fact}[theorem]{Fact}

\newtheorem{definition}[theorem]{Definition}

\newtheorem{problem}[theorem]{Problem}
\newtheorem{question}[theorem]{Question}
\crefname{question}{Question}{Questions}

\DeclarePairedDelimiter\floor{\lfloor}{\rfloor}
\DeclarePairedDelimiter\ket{\lvert}{\rangle}
\DeclarePairedDelimiter\bra{\langle}{\rvert}

\DeclareMathOperator*{\argmin}{arg\,min}
\newcommand{\E}{\mathop{\mathbb{E}\/}}

\newcommand{\tr} {\operatorname{tr}}

\newcommand{\rank} {\operatorname{rank}}

\newcommand{\ketbra}[2]{\ensuremath{\ket{#1}\!\bra{#2}}}

\newcommand{\kett}[1]{|#1\rangle\!\rangle}
\newcommand{\bbra}[1]{\langle\!\langle#1|}
\newcommand{\kettbbra}[2]{\ensuremath{\kett{#1}\!\bbra{#2}}}
\newcommand{\bbrakett}[2]{\ensuremath{\langle\!\langle{#1}\vert{#2}\rangle\!\rangle}}

\allowdisplaybreaks

\newcommand{\calC}{\mathcal{C}}

\newcommand{\calE}{\mathcal{E}}
\newcommand{\calF}{\mathcal{F}}
\newcommand{\calG}{\mathcal{G}}
\newcommand{\calH}{\mathcal{H}}

\newcommand{\calN}{\mathcal{N}}
\newcommand{\calO}{\mathcal{O}}

\newcommand{\calV}{\mathcal{V}}

\newcommand{\bb}{\boldsymbol{b}}

\newcommand*{\eps}{\varepsilon}

\newcommand{\qchannel}{\textbf{\textup{QChan}}}
\newcommand{\isochannel}{\textbf{\textup{ISO}}}

\renewcommand{\bb}{\begin{equation}\begin{aligned}\hspace{0pt}}
\newcommand{\ee}{\end{aligned}\end{equation}}

\newcommand*{\proj}[1]{\ketbra{#1}}

\usepackage{dsfont}
\usepackage{scalerel}

\newcommand{\Deltaout}{{\Delta_{\mathrm{out}}}}
\newcommand{\Deltain}{{\Delta_{\mathrm{in}}}}
\newcommand{\Proj}{\textup{Proj}}

\renewcommand{\Pr}{\mathbb{P}}

\begin{document}

\title{Quantum channel tomography: optimal bounds\\ in the near-boundary regime}
\date{}

\author{
Kean Chen\thanks{University of Pennsylvania, USA. Email: \texttt{keanchen.gan@gmail.com}}\and
Aadil Oufkir\thanks{University Mohammed VI Polytechnic, Morocco. Email: \texttt{aadil.oufkir@gmail.com}}
}

\maketitle

\begin{abstract}
Quantum channel tomography (also known as quantum process tomography) is a fundamental task in quantum information and quantum computing, aiming to learn the full description of an unknown quantum channel from black-box queries. For channels with input dimension $d_1$, output dimension $d_2$, and Kraus rank at most $r$, \cite{chen2026quantum} identifies the \textit{dilation rate} $\tau=rd_2/d_1\geq 1$ as a key parameter governing a Heisenberg-to-classical transition in query complexity. While the optimal bounds in the boundary regime ($\tau=1$) and the away-from-boundary regime ($\tau\geq 1+\Omega(1)$) are now known, the precise behavior in the near-boundary regime ($1<\tau<1+o(1)$) has remained unresolved.

In this paper, we close this gap, thereby fully settling the query complexity of quantum channel tomography.
Specifically, we show that the query complexity is
\begin{itemize}
    \item $\Omega(rd_1d_2/\varepsilon^2)$ for diamond norm error $\varepsilon$, matching the upper bound by \cite{mele2025optimallearningquantumchannels,chen2026quantum},
    \item $\Theta(d_1^2/\varepsilon+\min\{(\tau-1)d_1^2,(\tau-1)^2d_1^2d_2^2\}/\varepsilon^2)$ for   Choi trace norm error $\varepsilon$.
\end{itemize}
Our results reveal two qualitatively different forms of the Heisenberg-to-classical transition. For diamond norm tomography, the Heisenberg scaling disappears suddenly as soon as $\tau$ leaves the boundary regime. In contrast, Choi trace norm tomography exhibits a gradual transition from Heisenberg to classical scaling as $\tau$ moves from the boundary regime to the away-from-boundary regime.
\end{abstract}
\newpage

\section{Introduction}
\textit{Quantum channel tomography} (also known as \textit{quantum process tomography}) is a fundamental task in quantum information and quantum computing, aiming to reconstruct an unknown quantum channel from black-box queries. More specifically, given query access to an unknown quantum channel \(\mathcal{E}\), the goal is to learn a full classical description of \(\mathcal{E}\) to within a prescribed error \(\varepsilon\), with high probability. A central question is to determine the optimal number of queries required for this task. We consider channels with input dimension \(d_1\), output dimension \(d_2\), and Kraus rank at most \(r\), under two standard error metrics: the diamond norm and the   Choi trace norm, corresponding to the worst-case and average-case metrics, respectively.

One of the most intriguing aspects of quantum channel tomography is the possibility of achieving \textit{Heisenberg scaling} $1/\varepsilon$, in contrast to the \textit{classical scaling} $1/\varepsilon^2$. Understanding when such a quantum advantage is possible has been a central problem in quantum learning theory. The work~\cite{chen2026quantum} identifies the \textit{dilation rate}
\[
\tau:=\frac{rd_2}{d_1}\geq 1
\]
as a key parameter governing the query complexity of channel tomography, where the inequality is due to the trace preservation of quantum channels. In particular, it reveals a Heisenberg-to-classical phase transition between two parameter regimes:
\begin{itemize}
\item Boundary regime (\(\tau=1\)): The optimal query complexity is \(\Theta(rd_1d_2/\varepsilon)\) for both   Choi trace norm and diamond norm errors \cite{chen2026quantum,he2026query}. Thus, Heisenberg scaling is achievable in this regime.
\item Away-from-boundary regime (\(\tau\geq 1+\Omega(1)\)): The optimal query complexity becomes \(\Theta(rd_1d_2/\varepsilon^2)\) for both error metrics \cite{10.1145/2897518.2897544,mele2025optimallearningquantumchannels,chen2026quantum}, exhibiting classical scaling.
\end{itemize}

Although the Heisenberg-to-classical phase transition in quantum channel tomography has been identified in~\cite{chen2026quantum}, its fine-grained behavior in the near-boundary regime ($1<\tau<1+o(1)$) remains unclear. In this regime, the dilation rate $\tau$ can be arbitrarily close to the boundary value $1$. The work~\cite{chen2026quantum} establishes upper and lower bounds involving both Heisenberg and classical scaling terms, but these bounds do not match in their dependence on $\tau-1$. Consequently, the precise evolution of the optimal query complexity as $\tau$ moves away from the boundary remained unknown. In particular, it is unclear whether the classical-scaling contribution emerges abruptly or grows gradually as the gap $\tau-1$ increases.

A further subtlety is that the transition may depend on the choice of error metric. The diamond norm characterizes the worst-case distinguishability of quantum channels, whereas the   Choi trace norm measures the trace distance between their normalized Choi operators, characterizing the average-case distinguishability. Although both metrics behave the same in the boundary and away-from-boundary regimes, this does not necessarily imply that they behave the same in the near-boundary regime.

Therefore, a fine-grained characterization of the Heisenberg-to-classical transition is required to determine the precise dependence of the optimal query complexity.
In other words:
\begin{question}
How exactly does the Heisenberg-to-classical transition unfold in the near-boundary regime?
%What is the fine-grained behavior of the Heisenberg-to-classical transition in the near-boundary regime?
\end{question}

\subsection{Our results}
In this paper, we answer this question by establishing matching upper and lower bounds in the near-boundary regime, thereby completing the characterization of the query complexity of quantum channel tomography, for the entire parameter regime. Surprisingly, our results reveal two qualitatively different transitions. Under diamond norm, Heisenberg scaling disappears abruptly as soon as \(\tau>1\), even when \(\tau\) is arbitrarily close to \(1\), exhibiting an abrupt transition. In contrast, under   Choi trace norm, the query complexity exhibits a gradual transition, with the classical scaling contribution depending continuously on the distance from the boundary. 

Throughout the statements below, we consider feasible non-trivial parameters $d_1\ge 1$, $d_2\ge 2$, $1\le r \le d_1d_2 $ and $d_1\le rd_2$. All our results hold for success probability at least $2/3$. %and $0 < \varepsilon \le \eps_0$ where $\eps_0 = 10^{-12}$ is a universal non-optimized constant.

Our results on diamond norm tomography are shown as follows, and are also summarized in \cref{tab:diamond-bounds}.
\begin{theorem}[Diamond norm, \cref{thm:near-boundary-query-lower} restated]\label{thm-10310222}
Let $1<\tau=rd_2/d_1<4/3$. Tomography of an unknown quantum channel with input dimension $d_1$, output dimension $d_2$, Kraus rank at most $r$, and to within diamond norm error $\varepsilon$, has query complexity at least
\[\Omega\!\left(\frac{rd_1d_2}{\varepsilon^2}\right).\]
\end{theorem}

Combining \cref{thm-10310222} with the results from \cite{mele2025optimallearningquantumchannels,chen2026quantum,he2026query} (see \cref{tab:diamond-bounds}), we finally settle the query complexity of diamond norm channel tomography for the entire parameter regime.
\begin{corollary}[Diamond norm, entire regime]\label{coro-11010003}
Let $1\leq \tau=rd_2/d_1$. Tomography of an unknown quantum channel with input dimension $d_1$, output dimension $d_2$, Kraus rank at most $r$, and to within diamond norm error $\varepsilon$, has query complexity
\[\Theta\!\left(\frac{rd_1d_2}{\varepsilon^{2-\mathbbm{1}_{\tau=1}}}\right).\]
\end{corollary}
Notably, \cref{coro-11010003} reveals an abrupt Heisenberg-to-classical phase transition in diamond norm channel tomography: the optimal query complexity switches from Heisenberg scaling to classical scaling as soon as $\tau$ exceeds $1$. In particular, when $rd_2 = d_1+1$, the complexity is already $\Theta(d_1^2/\eps^2)$. 

\begin{table}[h]
\centering
\small
\begin{tabular}{|c|c|c|c|}
\hline
& \begin{tabular}[c]{@{}c@{}}\noalign{\smallskip} Boundary \\ $\tau=1$ \end{tabular}
& \begin{tabular}[c]{@{}c@{}}\noalign{\smallskip} Near-boundary \\ $1<\tau<1+c$ \end{tabular} 
&\begin{tabular}[c]{@{}c@{}}\noalign{\smallskip} Away-from-boundary \\ $\tau\geq 1+c$ \end{tabular} \\
\hline
Upper bound & \begin{tabular}[c]{@{}c@{}}\noalign{\smallskip} $O\!\left(\dfrac{rd_1d_2}{\varepsilon}\right)$ \\ \noalign{\smallskip} \cite{he2026query} \end{tabular}
& \multicolumn{2}{c|}{\begin{tabular}[c]{@{}c@{}}\noalign{\smallskip} $O\!\left(\dfrac{rd_1d_2}{\varepsilon^2}\right)$ \\ \noalign{\smallskip} \cite{mele2025optimallearningquantumchannels,chen2026quantum}\end{tabular}} \\ \hline
Lower bound & \begin{tabular}[c]{@{}c@{}}\noalign{\smallskip} $\Omega\!\left(\dfrac{rd_1d_2}{\varepsilon}\right)$ \\ \noalign{\smallskip} \cite{chen2026quantum}\end{tabular}
&\begin{tabular}[c]{@{}c@{}}\noalign{\smallskip} $\Omega\!\left(\dfrac{d_1^2}{\varepsilon}+\dfrac{(\tau-1)^2d_1^2}{\varepsilon^2}\right)$ \\ \noalign{\smallskip} \cite{chen2026quantum} \end{tabular} $\Rightarrow$ \begin{tabular}[c]{@{}c@{}}\noalign{\smallskip} $\Omega\!\left(\dfrac{rd_1d_2}{\varepsilon^2}\right)$ \\ \noalign{\smallskip} \cref{thm:near-boundary-query-lower} \end{tabular}
& \begin{tabular}[c]{@{}c@{}}\noalign{\smallskip} $\Omega\!\left(\dfrac{rd_1d_2}{\varepsilon^2}\right)$ \\ \noalign{\smallskip} \cite{chen2026quantum} \end{tabular} \\
\hline
\end{tabular}
\vspace{-2mm}
\caption{Diamond norm channel tomography. Here, $c\in(0,1/3]$ is an arbitrary but fixed constant. %UB: upper bound, LB: lower bound.
}
\label{tab:diamond-bounds}
\end{table}

Our results on   Choi trace norm tomography are shown as follows, and are also summarized in \cref{tab:choi-bounds}.
\begin{theorem}[Choi trace norm, \cref{thm-10300424,thm:near-boundary-query-lower} restated]\label{thm-10310227}
Let $1<\tau=rd_2/d_1<4/3$. Tomography of an unknown quantum channel with input dimension $d_1$, output dimension $d_2$, Kraus rank at most $r$, and to within   Choi trace norm error $\varepsilon$, has query complexity
\[\Theta\!\left(\frac{d_1^2}{\varepsilon}+\frac{\min\{(\tau-1)d_1^2,(\tau-1)^2d_1^2d_2^2\}}{\varepsilon^2}\right).\]
\end{theorem}

Combining \cref{thm-10310227} with the results from \cite{10.1145/2897518.2897544,chen2026quantum} (see \cref{tab:choi-bounds}), we finally settle the query complexity of   Choi trace norm channel tomography for the entire parameter regime.
\begin{corollary}[Choi trace norm, entire regime]\label{coro-11010006}
Let $1\leq \tau=rd_2/d_1$. Tomography of an unknown quantum channel with input dimension $d_1$, output dimension $d_2$, Kraus rank at most $r$, and to within   Choi trace norm error $\varepsilon$, has query complexity
\[\Theta\!\left(\frac{d_1^2}{\varepsilon}+\frac{\min\{(\tau-1)d_1^2,(\tau-1)^2d_1^2d_2^2\}}{\varepsilon^2}\right).\]
\end{corollary}
\begin{proof}
Note that $\Theta(\frac{d_1^2}{\varepsilon}+\frac{\min\{(\tau-1)d_1^2,(\tau-1)^2d_1^2d_2^2\}}{\varepsilon^2})$ becomes the optimal complexity $\Theta(\frac{rd_1d_2}{\varepsilon})$ and $\Theta(\frac{rd_1d_2}{\varepsilon^2})$ for $\tau=1$ and $\tau\geq 1+\Omega(1)$, respectively. Thus they can be written in a unified form.
\end{proof}
In contrast to the abrupt phase transition under the diamond norm in \cref{coro-11010003}, \cref{coro-11010006} reveals a gradual Heisenberg-to-classical transition under the   Choi trace norm, with the classical-scaling contribution varying continuously as $\tau$ moves away from the boundary.

\begin{table}[h]
\centering
\small
\setlength{\tabcolsep}{4pt}
\begin{tabular}{|c|c|c|c|}
\hline
& \begin{tabular}[c]{@{}c@{}}\noalign{\smallskip} Boundary \\  $\tau=1$ \end{tabular}
& \begin{tabular}[c]{@{}c@{}}\noalign{\smallskip} Near-boundary \\ $1<\tau<1+c$ \end{tabular} 
&\begin{tabular}[c]{@{}c@{}}\noalign{\smallskip} Away-from-boundary \\ $\tau\geq 1+c$ \end{tabular}  \\ \hline
UB &  \multirow{2}{*}{\begin{tabular}[c]{@{}c@{}}\noalign{\medskip}\noalign{\smallskip} $\Theta\!\left(\dfrac{rd_1d_2}{\varepsilon}\right)$ \\ \noalign{\smallskip} \cite{chen2026quantum}\end{tabular}} 
& \begin{tabular}[c]{@{}c@{}}\noalign{\smallskip} $O\!\left(\dfrac{d_1^2}{\varepsilon}+\dfrac{(\tau-1)d_1^2}{\varepsilon^2}\right)$ \\ \noalign{\smallskip} \cite{chen2026quantum} \end{tabular}\!\!\!
\multirow{2}{*}{\begin{tabular}[c]{@{}c@{}}\noalign{\bigskip}$\Rightarrow$ $\Theta\!\left(\dfrac{d_1^2}{\varepsilon}+\dfrac{\min\{(\tau-1)d_1^2,(\tau-1)^2d_1^2d_2^2\}}{\varepsilon^2}\right)$ \\ \noalign{\smallskip} \cref{thm-10300424,thm:near-boundary-query-lower} \end{tabular}}
& \begin{tabular}[c]{@{}c@{}}\noalign{\smallskip} $O\!\left(\dfrac{rd_1d_2}{\varepsilon^2}\right)$\\ \noalign{\smallskip} \cite{10.1145/2897518.2897544}\end{tabular} \\ \cline{1-1} \cline{4-4}
LB & & \begin{tabular}[c]{@{}c@{}} \noalign{\smallskip}\noalign{\smallskip} $\Omega\!\left(\dfrac{d_1^2}{\varepsilon} +\dfrac{(\tau-1)^2}{\varepsilon^2}\right)$ \\ \noalign{\smallskip} \cite{chen2026quantum} \end{tabular}\,\,
\hphantom{\begin{tabular}[c]{@{}c@{}} \noalign{\bigskip} $\rightarrow$ $\Theta\!\left( \dfrac{d_1^2}{\varepsilon} +\dfrac{\min\{1,(\tau-1)d_2^2\}(\tau-1)d_1^2} {\varepsilon^2}\right)$ \\ \noalign{\smallskip} \cref{thm-10300424,thm:near-boundary-query-lower} \end{tabular}}
& \begin{tabular}[c]{@{}c@{}}\noalign{\smallskip} $\Omega\!\left(\dfrac{rd_1d_2}{\varepsilon^2}\right)$\\ \noalign{\smallskip} \cite{chen2026quantum}\end{tabular} \\
\hline
\end{tabular}
\vspace{-2mm}
\caption{Choi trace norm channel tomography. Here, $c\in(0,1/3]$ is an arbitrary but fixed constant. UB: upper bound, LB: lower bound. The upper bound $O(\frac{rd_1d_2}{\varepsilon^2})$ is obtained by applying the state tomography algorithm~\cite{10.1145/2897518.2897544} to the Choi state.}
\label{tab:choi-bounds}
\end{table}

\subsection{Overview of techniques}
Our upper and lower bounds use the geometry of the Stinespring dilation $V : \mathbb{C}^{d_1} \to \mathbb{C}^{rd_2}$. Let $\alpha = rd_2-d_1$ and we assume that $0<\alpha<d_1$. We can see that the image of $V$ occupies all but $\alpha$ dimensions of the output space. 
%In the near-boundary regime, it is important to keep track of these remaining dimensions in order to determine the precise dependence on $\tau-1$.

\begin{figure}[t]
    \centering
    \includegraphics[width=0.95\linewidth]{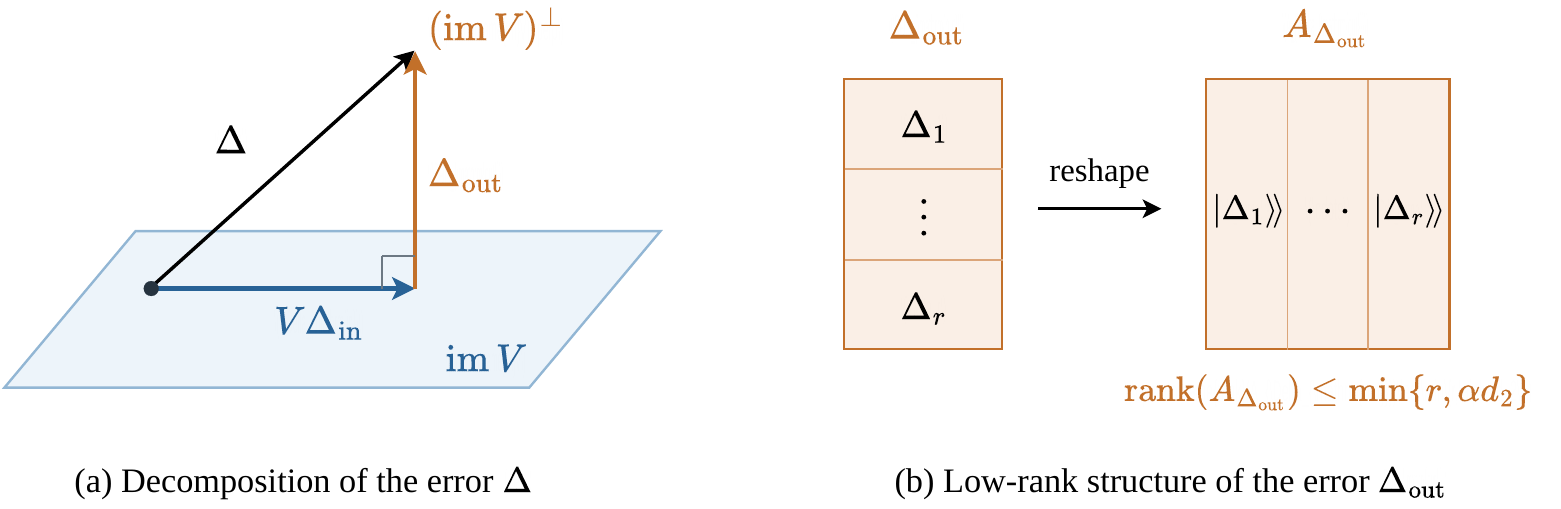}
    \caption{Two ingredients of the upper bound.
    (a) The error $\Delta$ splits into an in-image component
    $V\Deltain$ and an out-of-image component $\Deltaout$.
    (b) Kraus reshaping reveals the low-rank structure of
    $\Deltaout$, which sharpens the error bound after tracing out
    the ancilla.}
    \label{fig:upper-bound-techniques}
\end{figure}

\paragraph{Upper bound.}
By the parallel-query dilation reduction of~\cite[Theorem 1.5]{chen2026quantum},
we can estimate the Stinespring isometry
$V:\mathbb C^{d_1}\to\mathbb C^{d_2}\otimes\mathbb C^r$ and then trace out
the ancilla. To organize the estimation error, we first remove the
component of $\widehat V$ parallel to $V$ in the Frobenius inner product.
Since $\|V\|_{\mathrm F}^2=d_1$, this component is $zV$, with
$z\coloneqq\tr(V^\dagger\widehat V)/d_1$. The residual error
$\Delta\coloneqq\widehat V-zV$ therefore satisfies
$\tr(V^\dagger\Delta)=0$. As illustrated in
\cref{fig:upper-bound-techniques}(a), we further decompose it as
\[
\Delta=V\Deltain+\Deltaout,
\qquad \tr(\Deltain)=0,\qquad V^\dagger\Deltaout=0.
\]
Here $V\Deltain$ is the in-image error, while $\Deltaout$ is the out-of-image error that lies in the
$\alpha$-dimensional orthogonal complement of $\operatorname{im}V$.
Note that $\Deltain$ and $\Deltaout$ mirrors the Type I and Type II hard instances of~\cite{chen2026quantum}, which perturb an isometry within a fixed image and outside that image, respectively.
The scalar $z$ simply records the overall overlap: since both maps
are isometries, $1-|z|^2=\|\Delta\|_{\mathrm F}^2/d_1$.
Refining the covariant isometry estimator of~\cite{yoshida2025quantum}, we show
(see \cref{thm-10240443}) that
\begin{equation}\label{eq-11021343}
\E\|\Deltain\|_{\mathrm F}^2=O\!\left(\frac{d_1^5}{n^2}\right),
\qquad
\E\|\Deltaout\|_{\mathrm F}^2=O\!\left(\frac{\alpha d_1^2}{n}\right).
\end{equation}
Thus the in-image error enjoys Heisenberg scaling, while the
out-of-image error has classical scaling but is suppressed by $\alpha$.

Another key step is to exploit the low-rank structure
of $\Deltaout$ after reshaping the Kraus operators, as shown in
\cref{fig:upper-bound-techniques}(b). Since $\Deltaout$ takes values
in an $\alpha$-dimensional subspace, its reshaped matrix satisfies
$\rank(A_{\Deltaout})\leq\beta\coloneqq\min\{r,\alpha d_2\}$.
Consequently, the cross term arising after tracing out the ancilla obeys
\[
\|A_VA_{\Deltaout}^\dagger\|_1
\leq \sqrt{d_2\beta}\,\|\Deltaout\|_{\mathrm F},
\]
where we use $\|A_V\|_{\mathrm{op}}\leq\sqrt{d_2}$. Thus $\|C_{\widehat{\mathcal{E}}}-C_{\mathcal{E}}\|_1$ can be well-bounded (see \cref{lemma-10242227}), where $\mathcal{E}$ is the channel obtained from $\mathcal{V}$ after tracing out the ancilla. Combining this with the separate error estimates in \cref{eq-11021343} gives
\[
\frac{1}{d_1}\E\left\|C_{\widehat{\mathcal E}}-C_{\mathcal E}\right\|_1
=O\!\left(
\frac{d_1^2}{n}
+\sqrt{\frac{\min\{\alpha d_1,\alpha^2d_2^2\}}{n}}
\right),
\]
which yields the claimed
$O\!\left(d_1^2/\varepsilon+
\min\{\alpha d_1,\alpha^2d_2^2\}/\varepsilon^2\right)$
query upper bound. 

\paragraph{Lower bound.} The proofs of the lower bounds use a two stage approach as in \cite{chen2026quantum}. First, we construct packing nets that have desirable properties. Then, we show a hardness result for a channel discrimination problem. Finally, we combine the constructed packing nets with the hardness of channel discrimination problem to deduce our lower bounds. 

The construction of packing nets away from the boundary in \cite{chen2026quantum} uses a small perturbation whose image is orthogonal to that of a fixed center map. Its discrimination proof groups tensor-product terms by the number of perturbation components. The orthogonality feature is the key of the classical scaling. However, near the boundary, it is not clear how to fit the perturbation into the available output space while respecting the orthogonality condition and preserving channel separation. In particular, the resulting diamond-norm lower bound of \cite{chen2026quantum} has classical term $(\tau-1)^2d_1^2/\eps^2 \underset{ \tau \to 1}{\rightarrow}0$. In this work, we replace the small-perturbation family by a new family whose changing components are isometries of input dimensions $d' = \floor{d_1/2}$. 

\begin{figure}[!t]
\centering
\begin{tikzpicture}[
 x=0.97cm,y=0.78cm,font=\small,
 >={Stealth[length=2.2mm]},
 reference/.style={draw=teal!75!black,fill=teal!14,line width=0.9pt},
 signal/.style={draw=blue!75!black,fill=blue!18,line width=0.9pt},
 ambient/.style={draw=Orange!80!black,fill=Orange!9,line width=0.9pt},
 dimension/.style={decorate,decoration={brace,amplitude=4pt},line width=0.6pt}
]
 %\node[font=\bfseries] at (7.4,7.25) {The construction in $d'$ notation};
 \node[font=\small\bfseries] at (3.65,6.65) {Input: dimension $d_1$};
 \node[font=\small\bfseries] at (9.65,6.65) {Dilation output: dimension $rd_2$};
 % Fixed reference map: the two narrow boxes have equal dimension.
 \node[anchor=east,align=right,text width=2.7cm] at (2.65,6.05)
   {Reference input};
 \draw[reference] (3.35,4.4) rectangle (3.95,6.1);
 \draw[dimension,teal!75!black] (3.12,4.4) -- (3.12,6.1)
   node[midway,left=6pt] {$d_1-d'$};
 \draw[->,teal!75!black,line width=1.1pt] (4.2,5.25) -- (9.05,5.25)
   node[midway,above=4pt] {$F$}
   node[midway,below=4pt,font=\footnotesize] {fixed isometry};
 \draw[reference] (9.35,4.4) rectangle (9.95,6.1);
 \draw[dimension,teal!75!black] (10.18,6.1) -- (10.18,4.4)
   node[midway,right=6pt] {$d_1-d'$};
 \node[anchor=west,align=left,text width=3.5cm] at (11.7,5.25)
   {Fixed reference image\\$\operatorname{im}F$};
 % Random signal range is drawn inside the whole complementary output space.
 \node[anchor=east,align=right,text width=2.7cm] at (2.65,2.9)
   {Signal input};
 \draw[signal] (3.35,0.75) rectangle (3.95,2.85);
 \draw[dimension,blue!75!black] (3.12,0.75) -- (3.12,2.85)
   node[midway,left=6pt] {$d'$};
 \draw[ambient] (8.95,-0.15) rectangle (10.35,3.65);
 \draw[signal] (9.35,0.75) rectangle (9.95,2.85);
 \draw[->,blue!75!black,line width=1.1pt]
   (4.2,1.8) to[out=0,in=180] (9.32,1.8);
 \node[align=center,text width=3.8cm,text=blue!75!black]
   at (6.55,2.55) {$W_PU$};
 \node[align=center,text width=3.9cm,font=\footnotesize]
   at (6.55,0.85) {Random range $P$;\\Haar mask $U\in\mathbb U_{d'}$};
 % Separate braces measure the blue inner box and the orange outer box.
 \draw[dimension,blue!75!black] (10.1,2.85) -- (10.1,0.75)
   node[midway,right=6pt,fill=Orange!9,inner sep=1pt] {$d'$};
 \draw[dimension,Orange!80!black] (11.0,3.65) -- (11.0,-0.15)
   node[midway,right=6pt,fill=white,inner sep=2pt] {$d'+\alpha$};
 \node[anchor=west,align=left,text width=3.5cm,text=blue!75!black]
   at (11.7,2.95) {Signal range\\$\operatorname{im}P$};
 \node[anchor=west,align=left,text width=3.5cm,text=Orange!80!black]
   at (11.7,0.1) {Outer space $\operatorname{im}\Pi$\\$\Pi=I_{rd_2}-FF^\dag$};
 \node[align=center,font=\footnotesize,text width=6.5cm]
   at (9.65,-0.85)
   {$\dim\operatorname{im}(\Pi-P)=\alpha=rd_2-d_1$};
\end{tikzpicture}
\caption{The hard family $V_{P,U} = \begin{pmatrix}
    F & W_PU
\end{pmatrix}$. The reference map $F$ is fixed, the signal range $P$ is the unknown label, and a Haar-random $U$ hides the basis within that range.  }
\label{fig:range-construction-general-intro}
\end{figure}
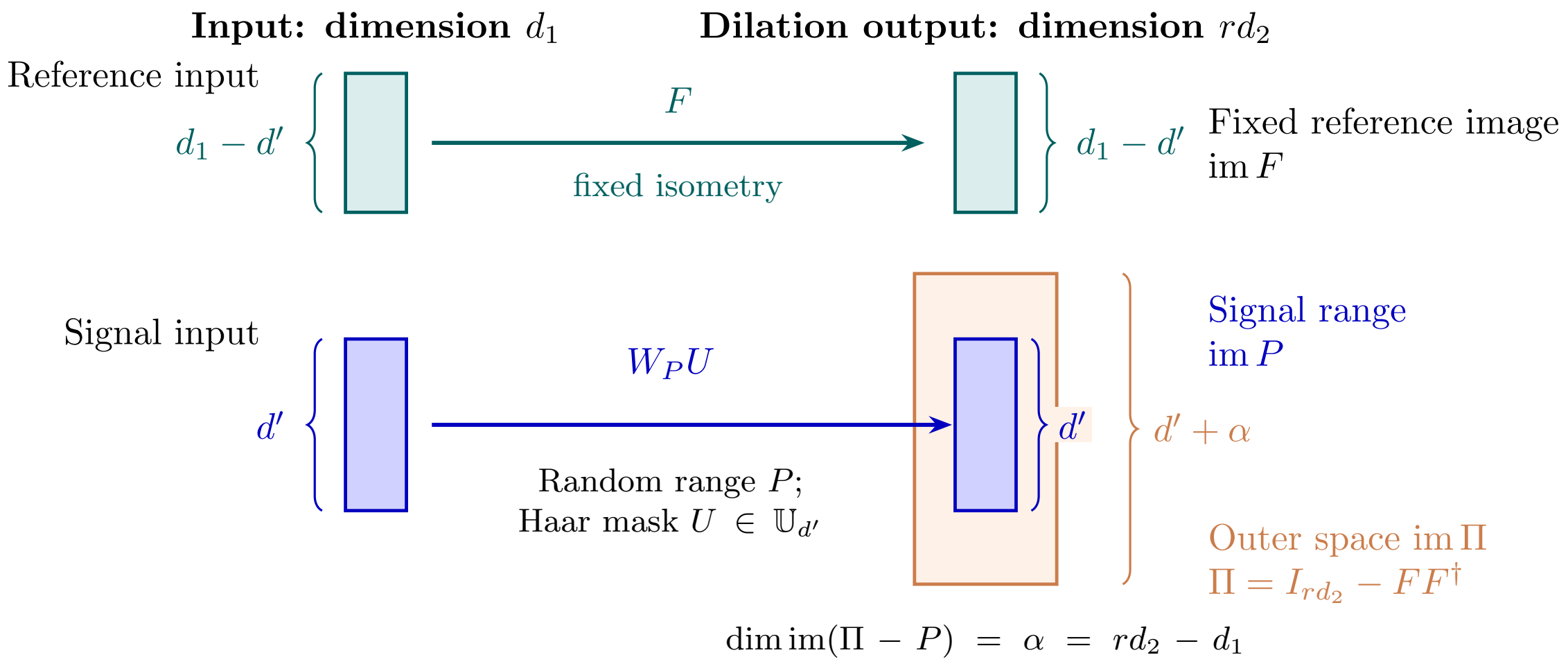

Specifically, we split the input into a fixed reference part and a signal part of dimension $d'$, and define the dilation and the corresponding channel as 
\[V_{P,U} = \begin{pmatrix}
    F & W_PU
\end{pmatrix}, \quad  \calV_{P,U} = V_{P,U}(\cdot) V_{P,U}^\dagger  , \quad  \calE_{P,U} \coloneqq \tr_{\rm anc}\circ \calV_{P,U}, \]
where $F$ is the (fixed) reference isometry, $P$ is a projector of rank $d'$ in a space of dimension $d'+ \alpha$, $W_P$ is an isometry onto that range, and $U$ is an independent Haar random unitary. The unknown label is the projector $P$ and the unitary $U$ hides the choice of basis within its range. See  \cref{fig:range-construction-general-intro} for an illustration.   The reference preserves information about the signal through the off-diagonal blocks of the channel's Choi operator. Crucially, we prove that different ranges give separated channels uniformly over all choices of the masks $U$, so the extra randomness makes discrimination harder without destroying the information that tomography must recover.

As in \cite{chen2026quantum}, we consider a problem of discriminating between quantum channels and we show a hardness result on the query complexity. Our proof refines a Haar-average comparison method that was developed in \cite{hashimoto2010unitary,bavaresco2022unitary} and used  by \cite{haah2023query} to bound the success probability of identifying one of $M$ unitary channels in dimension $D$ by $\frac{1}{M} \binom{n+D^2-1}{n}$ where $n$ is the query complexity. Our new contribution is a comparison for masked isometries that makes the dependence on $\alpha$ explicit. Let 
$\mathcal{N}$ be a set of quantum channels that we want to discriminate (later, this will be our packing net). Let 
\[\Gamma_P = \mathbb{E}_U C_{\calV_{P,U}}^{\otimes n}, \quad \Gamma =\mathbb{E}_P \Gamma_P.\]
$\Gamma$ is a valid channel comb so every normalized tester $\{T_P\}_P$ satisfies $\left(\sum_P T_P\right) \star \Gamma=1$. Thus an operator inequality $\Gamma_P\preceq L_n\Gamma$ for every $P$ immediately gives
\begin{equation}\label{eq:success-UB}
    \Pr[\textup{success}]
=\frac1{|\mathcal N|}\sum_{P\in\mathcal N}T_P\star\Gamma_P
\leq\frac{L_n}{|\mathcal N|}.
\end{equation}
This is the standard comparison step.  Our goal is therefore to obtain a bound on $L_n$ that reflects the freedom to choose the signal range, while removing the freedom to choose a basis within that range. 

We consider first averaging over  $U$. Averaging its global phase  removes cross terms between different numbers $\ell$ of signal uses. For each $\ell$, Schur--Weyl duality decomposes the signal into blocks indexed by $\lambda\vdash_{d'}\ell$. Let
$$
a_\lambda=\dim\mathcal Q_\lambda^{d'},
\qquad
b_\lambda=\dim\mathcal Q_\lambda^{d'+\alpha}.
$$
After averaging over the basis, the dependence on $P$ in the $\lambda$-block is through a projector $\Pi_{\lambda,P}$ of rank $a_\lambda$ on a $b_\lambda$-dimensional space. Averaging the range itself makes this projector uniform:
$$
\mathbb E_P\Pi_{\lambda,P}
=\frac{a_\lambda}{b_\lambda}I.
$$
Thus for every $P$, 
$$
\Pi_{\lambda,P}
\preceq
\frac{b_\lambda}{a_\lambda}\, 
\mathbb E_Q\Pi_{\lambda,Q}.
$$
Hence, it suffices to take
$$
L_n=
\max_{\substack{0\leq\ell\leq n\
\lambda\vdash_{d'}\ell}}
\frac{\dim\mathcal Q_\lambda^{d'+\alpha}}
{\dim\mathcal Q_\lambda^{d'}}=\max_{\substack{0\leq\ell\leq n\
\lambda\vdash_{d'}\ell}}
\prod_{i=1}^{d'}\prod_{j=1}^{\alpha}
\frac{\lambda_i+d'+j-i}{d'+j-i}
\leq
\left(e+\frac{2e n}{d'(d'+\alpha)}\right)^{\alpha d'}.
$$
Finally,
$$
\Gamma_P\preceq
\left(e+\frac{2en}{d'(d'+\alpha)}\right)^{\alpha d'}\Gamma,
\qquad
\Pr[\mathrm{success}]
\leq
\frac{1}{|\mathcal N|}
\left(e+\frac{2en}{d'(d'+\alpha)}\right)^{\alpha d'},
$$
\cref{thm:hard-isom-disc}.

We now turn to the construction of the packing with a sufficiently large cardinality. Unlike the small-perturbation construction in~\cite{chen2026quantum}, the precision parameter enters through the distance between  the ranges. A Haar block law~\cite{fyodorov2007colour} and matrix-ball volume estimates~\cite{hua1963harmonic}, combined with trace-norm duality and the Blaschke--Santal\'o inequality~\cite{bianchi2015fourier}, give for a fixed projector $P$
$$
\mu\left\{Q:\|P-Q\|_1\leq
\frac{4\alpha d'}{e(d'+\alpha)}\eta\right\}
\leq\eta^{2\alpha d'}
$$
for $0<\eta<1$ (\cref{lem:haar-small-ball}). 
A maximal packing in a good set of Haar measure at least $1/2$ therefore has $|\mathcal N|\geq\tfrac12\eta^{-2\alpha d'}$.  Combining the two bounds gives
$$
\frac23\leq\Pr[\textup{success}]
\leq2\left[\eta^2\left(e+\frac{2en}{d'(d'+\alpha)}\right)\right]^{\alpha d'},
$$
hence $n=\Omega\!\left(\frac{d'(d'+\alpha)}{\eta^2}\right)$ for sufficiently small universal $\eta$. The same exponent $\alpha d'$ appears on both components of the upper bound in \cref{eq:success-UB}, so it cancels when we solve for $n$. Even when $\alpha=1$, the query complexity is still $\Theta(d_1^2/\eta^2)$.

It remains to make the range separation visible at the channel level. We choose $F$ so that its positive Choi eigenvalues are  bounded below, using a balanced reference construction from~\cite{chenoufkir2026incoherent}. For diamond norm, an input superposing the reference and signal parts detects a displacement of the signal range (\cref{lem:unif-sep-dia}). The resulting channel separation is $\Omega(\eta)$ and yields $\Omega(d_1^2/\varepsilon^2)$ queries even when $\alpha=1$.

For   Choi trace norm, the maximally entangled input averages over the entire input space. We additionally restrict to signal ranges whose orthogonal complements have a bounded environment marginal. A Gaussian argument shows that these ranges occupy at least half of Haar measure, so this restriction preserves the packing size up to a constant. A trace-norm witness in the reference--signal blocks then gives separation $\Omega(\sqrt \chi\,\eta/d_1)$, where $\chi=\min\{\alpha d_1,\alpha^2d_2^2\}$ (\cref{lem:unif-sep-choi,thm:packing}). The same discrimination bound therefore gives $\Omega(\chi/\varepsilon^2)$. Combining it with the Heisenberg lower bound completes the result. These two separation arguments explain why the same hard family gives an abrupt transition for diamond norm and a gradual transition for   Choi trace norm.

\subsection{Related work}
\label{sec:related-work}

\paragraph{Quantum state tomography.}

The study of quantum state tomography dates back to early work on optimal pure-state estimation~\cite{hayashi1998asymptotic,bruss1999optimal}. For general mixed states, Haah, Harrow, Ji, Wu, and Yu~\cite{Haah_2017} established nearly sample-optimal bounds, while O'Donnell and Wright~\cite{10.1145/2897518.2897544} developed an $O(rd/\varepsilon^2)$ upper bound for rank-$r$ state tomography in trace norm. Gu{\c{t}}{\u{a}}, Kahn, Kueng, and Tropp~\cite{GKKT20} analyzed projected least-squares tomography with nonasymptotic error guarantees.
On the lower-bound side, Scharnhorst, Spilecki, and Wright~\cite{scharnhorst2025optimal} recently established the matching $\Omega(rd/\varepsilon^2)$ lower bound for trace-norm tomography. More recently, Pelecanos, Spilecki, Tang, and Wright~\cite{pelecanos2025mixedstatetomographyreduces} reduced mixed-state tomography to pure-state tomography via random purification, yielding sample-optimal and gate-efficient state tomography.
Applying state tomography to a channel's normalized Choi state gives an $O(rd_1d_2/\varepsilon^2)$ upper bound for Choi trace-norm channel tomography, but does not capture the advantages of coherent channel queries near the boundary.

\paragraph{Unitary and isometry tomography.} Heisenberg scaling can be achieved for
Unitary estimation~\cite{yang2020optimal}. Haah, Kothari, O'Donnell, and Tang~\cite{haah2023query} established the optimal $\Theta(d^2/\varepsilon)$ query complexity for diamond-norm unitary tomography using sequential queries. For parallel queries, He, Li, and Yang~\cite{he2026query} achieved the optimal $O(d^2/\varepsilon)$ bound (see also \cite{scully2026nearoptimal} for a near-optimal bound). This yields the optimal diamond-norm upper bound for channel tomography in the boundary regime through the parallel dilation reduction~\cite{chen2025quantum,Girardi2025Dec,yoshida2025random}. Related work studies unitary-channel shadow estimation~\cite{he2026optimal}, which has a different learning objective. For isometry channels, Yoshida, Miyazaki, and Murao~\cite{yoshida2025quantum} developed an estimation framework under a fidelity loss, which we refine by separately analyzing errors inside and outside the image of the unknown isometry.

\paragraph{Quantum channel tomography.}
Both state tomography and unitary/isometry tomography are special cases of quantum channel tomography, corresponding to one-dimensional input ($d_1=1$) and Kraus rank one ($r=1$), respectively. Early works introduced channel tomography and its connection to the Choi representation~\cite{chuang1997prescription,poyatos1997complete,leung2000towards,d2001quantum}.
More recently, Mele and Bittel~\cite{mele2025optimallearningquantumchannels} and Chen, Yu, and Zhang~\cite{chen2025quantum} concurrently and independently established an $O(rd_1d_2/\varepsilon^2)$ upper bound under diamond norm error. The former analyzes a Choi-state estimator directly in diamond norm, while the latter develops a local-test framework that reduces channel learning to estimation of a Stinespring isometry without increasing the number of parallel queries. The latter work also showed that Heisenberg scaling is achievable in the boundary regime $rd_2=d_1$, even for non-unitary channels, obtaining $O(d_1^{5/2}/\varepsilon)$ and $O(d_1^2/\varepsilon)$ upper bounds under diamond norm and   Choi trace norm errors, respectively.

On the lower-bound side, a near-optimal Heisenberg-scaling lower bound in the boundary regime and an optimal classical-scaling lower bound in the away-from-boundary regime $rd_2\geq 2d_1$ were proved by Oufkir and Girardi~\cite{oufkir2026improved} and Chen, Zhang, and Yu~\cite{chen2026optimal}, respectively. The three works~\cite{chen2025quantum,oufkir2026improved,chen2026optimal} were subsequently subsumed and extended in \cite{chen2026quantum}, which identified the dilation rate $\tau=rd_2/d_1$ as governing a Heisenberg-to-classical phase transition. They obtained optimal bounds at the boundary for   Choi trace norm and away from the boundary for both metrics, while leaving the near-boundary regime unresolved. Our work closes this gap and characterizes the fine-grained transition under both error metrics.

Under the additional restriction of incoherent access, Surawy-Stepney, Kahn, Kueng, and Guta~\cite{surawy2022projected} studied projected least-squares estimators, and Oufkir~\cite{Oufkir_2023} established near-optimal bounds for nonadaptive protocols. More recently, Bravo-Prieto, Gong, and Mele~\cite{bravoprieto2026memory} proved an optimal $\Theta(d_1^3d_2^3/\varepsilon^2)$ bound without a rank promise, while Chen and Oufkir~\cite{chenoufkir2026incoherent} obtained the nearly optimal bound $\widetilde{\Theta}(d_1d_2r^2/\varepsilon^2)$ for channels of bounded Kraus rank $r$.

\subsection{Discussion}
Together with previous work, our results finally settle the optimal query complexity
of quantum channel tomography under both diamond norm and Choi trace norm,
across the entire parameter regime (see \cref{coro-11010003,coro-11010006}).
Our results also reveal two qualitatively different Heisenberg-to-classical
transitions: an abrupt transition under the diamond norm, where classical
scaling appears as soon as $\tau>1$, and a gradual transition under the
Choi trace norm, where the classical-scaling contribution grows as $\tau$
moves away from the boundary.

Another important observation is that adaptivity provides no asymptotic
query advantage in either setting. Indeed, our upper bounds are achieved
using parallel queries, while the matching lower bounds hold even for
arbitrary sequential, adaptive, and coherent algorithms. Thus, the optimal query
complexity can always be attained without adaptivity (sequential queries).

Several directions remain open. One is to develop computationally efficient
implementations of the query-optimal algorithms, including efficient
state preparation, measurements, and classical postprocessing.
Another is to understand channel tomography under quantum memory
constraints. In particular, it would be interesting to characterize
the tradeoff between query complexity and quantum memory. Existing results for incoherent
access~\cite{chenoufkir2026incoherent} provide a starting point.

\section{Preliminaries}\label{sec-770122}

\subsection{Notation}
\paragraph{Linear algebra.}
Given two orthonormal bases for Hilbert spaces $\mathcal{H}_0$ and $\mathcal{H}_1$, respectively, we can represent every linear operator $X:\mathcal{H}_0\rightarrow\mathcal{H}_1$ by a $\dim(\mathcal{H}_1)\times \dim(\mathcal{H}_0)$ matrix, and, for such a matrix \(X\), we use \(\kett{X}\in \mathcal{H}_1\otimes\mathcal{H}_0\) to denote the vector obtained by flattening the matrix $X$. It is easy to see the following properties:
\[\kett{\ketbra{\psi}{\phi}}=\ket{\psi}\ket{\phi^*}, \quad\quad\quad \kett{XYZ}=X\otimes Z^\textup{T} \kett{Y},\]
where \(\ket{\phi^*}\) denotes the entry-wise complex conjugate of \(\ket{\phi}\)  and $Z^\textup{T}$ denotes the transpose of the matrix $Z$ with respect to a given orthonormal basis. 
The inner product can be rewritten as \(\bbrakett{X}{Y}=\tr(X^\dag Y)\).
For two  Hermitian operators $X,Y$, we write $X\preceq Y$ to denote that $Y-X$ is positive semidefinite.

Let $n, m, d$ be positive integers such that $n\geq m$. Let $\mathcal{H}_1\cong\cdots\cong \mathcal{H}_n\cong\mathbb{C}^{d}$ be $n$ copies of a $d$-dimensional Hilbert space $\mathcal{H}$. 
Let $S\subseteq [n]=\{1,2,\ldots,n\}$ be a set of integers and let $\ket{\psi}\in\mathbb{C}^d$ be a state. We denote by
\[\ket{\psi}^{\otimes S}\]
the state $\ket{\psi}^{\otimes |S|}\in\bigotimes_{i\in S} \mathcal{H}_i$.
Therefore, if $\ket{\varphi}\in \mathbb{C}^d$ is another state, then we call
\[\ket{\psi}^{\otimes S} \otimes\ket{\varphi}^{\otimes [n]\setminus S}\]
the state $\bigotimes_{i=1}^n \ket{x_i}\in\bigotimes_{i=1}^n \mathcal{H}_i$, where $\ket{x_i}=\ket{\psi}$ for $i\in S$, and $\ket{x_i}=\ket{\varphi}$ for $i\notin S$.

\paragraph{Quantum channels and isometries.} We take $d_2\geq2$ throughout: when $d_2=1$, the only channel is the trace map, which requires no queries to learn. 
We denote by $\qchannel_{d_1,d_2}^r$ the set of all quantum channels $\mathcal{E}:\mathcal{L}(\mathbb{C}^{d_1})\rightarrow \mathcal{L}(\mathbb{C}^{d_2})$ having Kraus rank at most $r$.
We denote by $\isochannel_{d,D}$ the set of all isometries with input dimension $d$ and output dimension $D$ where $D\geq d$.

Any channel $\mathcal{E}\in\qchannel_{d_1,d_2}^r$ can be identified by its Choi-Jamio{\l}kowski operator~\cite{choi1975completely,jamiolkowski1972linear} as follows:
\begin{equation*}
    C_\mathcal{E}=(\mathcal{E}\otimes \mathcal{I})(\kettbbra{I}{I})\in\mathcal{L}(\mathbb{C}^{d_2}\otimes\mathbb{C}^{d_1}),
\end{equation*}
where we have denoted by $\kett{I}=\sum\limits_{i}\ket{i}\ket{i}\in \mathbb{C}^{d_1}\otimes \mathbb{C}^{d_1}$ an unnormalized maximally entangled state. We may simply call $C_\mathcal{E}$ Choi operator or (unnormalized) Choi state. 
Note that it is possible to write the Choi operator as $C_\mathcal{E}=\sum_{i=1}^r\kettbbra{E_i}{E_i}$, where $E_i$ are orthogonal Kraus operators and thus $\kett{E_i}$ are pairwise orthogonal vectors. Hence, the Kraus rank is equal to the rank of the Choi operator. As a consequence, one can easily see that $r\leq d_1d_2$.

\paragraph{Error metrics and query access.}
We use the    Choi trace norm distance $d_1^{-1}\|C_{\mathcal E}-C_{\mathcal F}\|_1$ and the diamond norm distance $\|\mathcal E-\mathcal F\|_\diamond$. For two channels with input space $\mathcal H_{\mathrm A}\cong\mathbb C^{d_1}$, the latter is defined by
$$
\|\mathcal E-\mathcal F\|_\diamond
=\max_{\substack{\rho_{\mathrm{AR}}\succeq0\\\tr\rho_{\mathrm{AR}}=1}}
\left\|\bigl((\mathcal E-\mathcal F)\otimes\mathcal I_{\mathrm R}\bigr)(\rho_{\mathrm{AR}})\right\|_1,
\qquad \mathcal H_{\mathrm R}\cong\mathbb C^{d_1},
$$
where $\mathcal I_{\mathrm R}$ is the identity channel on the reference system~\cite{watrous2018theory}.  In particular,
$$
\frac{1}{d_1}\|C_{\mathcal E}-C_{\mathcal F}\|_1\leq\|\mathcal E-\mathcal F\|_\diamond.
$$
An algorithm may use a reference system and quantum memory, and may choose its inputs adaptively between channel queries. Our lower bounds hold in this model; the upper bound in \cref{thm-10300424} uses parallel queries.
\paragraph{Kraus-vector reshaping.}
For a linear map $T:\mathbb C^s\to\mathcal H_{\mathrm B}\otimes\mathcal H_{\mathrm{anc}}$, where $\mathcal H_{\mathrm B}\cong\mathbb C^{d_2}$ and $\mathcal H_{\mathrm{anc}}\cong\mathbb C^r$, write $T=\sum_{e=1}^rT_e\otimes\ket e_{\mathrm{anc}}$. Define
\begin{equation}\label{eq:Kraus}
A_T=\sum_{e=1}^r\kett{T_e}\bra e_{\mathrm{anc}}:
\mathcal H_{\mathrm{anc}}\to\mathcal H_{\mathrm B}\otimes\mathbb C^s.
\end{equation}
For $T,S$ with the same domain and codomain, vectorization gives
\begin{align}
A_T^\dagger A_T&=\bigl(\tr_{\mathrm B}TT^\dagger\bigr)^{\mathrm T},&
A_TA_S^\dagger&=\tr_{\mathrm{anc}}\kettbbra T S,
\label{eq:reshaping-identities}\\
\tr(A_T^\dagger A_S)&=\tr(T^\dagger S),&
\|A_T\|_{\mathrm F}&=\|T\|_{\mathrm F}.
\label{eq:reshaping-inner-product}
\end{align}
In the partial-trace identity, tensor factors are placed in the order $\mathcal H_{\mathrm B}\otimes\mathbb C^s\otimes\mathcal H_{\mathrm{anc}}$. In particular, for an isometry $T$ inducing the channel $\mathcal E_T=\tr_{\mathrm{anc}}\circ\mathcal T$, we have
\begin{equation}\label{eq:Kraus-Choi}
C_{\mathcal E_T}=A_TA_T^\dagger,\qquad
\|A_T\|_{\mathrm{op}}\leq\sqrt{d_2}.
\end{equation}
The norm bound follows from $TT^\dagger\preceq I$ and the first identity in \cref{eq:reshaping-identities}. More generally,
\begin{equation}\label{eq:reshaping-rank}
\rank A_T\leq\min\{r,d_2\rank T\}.
\end{equation}
Indeed, the environment components of a basis for $\operatorname{im}T$ span a space of dimension at most $d_2\rank T$; the first identity in \cref{eq:reshaping-identities} then gives the rank bound.

\paragraph{Haar measures.}
We use $U\sim \mathbb{U}_d$ to denote that $U$ is sampled from the Haar distribution on $\mathbb{U}_d$.
We use $V\sim\isochannel_{d,D}$ to denote that $V$ is sampled from the Haar distribution on $\isochannel_{d,D}$.
Here the Haar distribution on $\isochannel_{d,D}$ is defined by picking an arbitrary $V_0\in\isochannel_{d,D}$ and set $V=UV_0$ for $U\sim\mathbb{U}_D$. One can easily check that this is well-defined since the distribution of $V$ does not depend on the choice of $V_0$.

\subsection{Quantum combs and testers}\label{sec-6290132}
\paragraph{Quantum combs.}
The framework of quantum combs~\cite{chiribella2008quantum,chiribella2009theoretical} provides a powerful description of higher-order transformations of quantum processes. More specifically, the Choi-Jamio{\l}kowski representation of quantum channels, which describes transformations of quantum states, can be extended to the higher-order setting of transformations of quantum processes. Such higher-order objects are called \textit{quantum combs}.

\begin{definition}[Quantum comb~\cite{chiribella2009theoretical}]\label{def-681501}
Let $n\geq 1$ be an integer. A deterministic $n$-comb, defined on a sequence of $2n$ Hilbert spaces $(\mathcal{H}_0,\mathcal{H}_1,\ldots,\mathcal{H}_{2n-1})$, is a positive semidefinite operator $X$ on $\bigotimes_{j=0}^{2n-1} \mathcal{H}_{j}$ such that there exists a sequence of operators $X^{(n)}, X^{(n-1)},\ldots, X^{(1)}, X^{(0)}$ satisfying
\begin{equation}\label{eq-681546}
\begin{split}
\tr_{\mathcal{H}_{2j-1}}\!\left(X^{(j)}\right)&=I_{\mathcal{H}_{2j-2}}\otimes X^{(j-1)},\quad 1\leq j \leq n,  
\end{split}
\end{equation}
where $X^{(n)}=X$ and $X^{(0)}=1$.
\end{definition}
It is easy to see that any convex combination of quantum $n$-combs is also a quantum $n$-comb.
Now, let us introduce the link product ``$\star$''.
\begin{definition}[Link product ``$\star$''~\cite{chiribella2008quantum,chiribella2009theoretical}]
\label{def-720255}
Let $X$ be a linear operator on $\mathcal{H}_{\bm{i}}=\mathcal{H}_{i_1}\otimes\mathcal{H}_{i_2}\otimes\cdots\otimes\mathcal{H}_{i_{n}}$ and let $Y$ be a linear operator on $\mathcal{H}_{\bm{j}}=\mathcal{H}_{j_1}\otimes\mathcal{H}_{j_2}\otimes\cdots\otimes\mathcal{H}_{j_{m}}$,
where $\bm{i}=(i_1,\ldots,i_n)$ is a sequence of pairwise distinct indices, and likewise for $\bm{j}=(j_1,\ldots,j_m)$.
Let $\bm{a}=\bm{i}\cap\bm{j}$ be the set of indices appearing in both $\bm{i}$ and $\bm{j}$ and $\bm{b}=\bm{i}\cup\bm{j}$ be the set of indices appearing in either $\bm{i}$ or $\bm{j}$.
Then, the combination of $X$ and $Y$ is
\[X\star Y\coloneqq \tr_{\mathcal{H}_{\bm{a}}}\!\left(X^{\textup{T}_{\mathcal{H}_{\bm{a}}}} \cdot Y\right)=\tr_{\mathcal{H}_{\bm{a}}}\!\left(X\cdot Y^{\textup{T}_{\mathcal{H}_{{\bm{a}}}}}\right),\]
where $\mathcal{H}_{\bm{a}}$ denotes the tensor product of subsystems labeled by the indices in $\bm{a}$, $\textup{T}_{\mathcal{H}_{\bm{a}}}$ denotes the partial transpose on $\mathcal{H}_{\bm{a}}$, both $X$ and $Y$ are treated as linear operators on $\mathcal{H}_{\bm{b}}$, extended by tensoring with the identity operator as needed.
\end{definition}

\begin{figure}[t]
    \centering
    \includegraphics[width=0.6\linewidth]{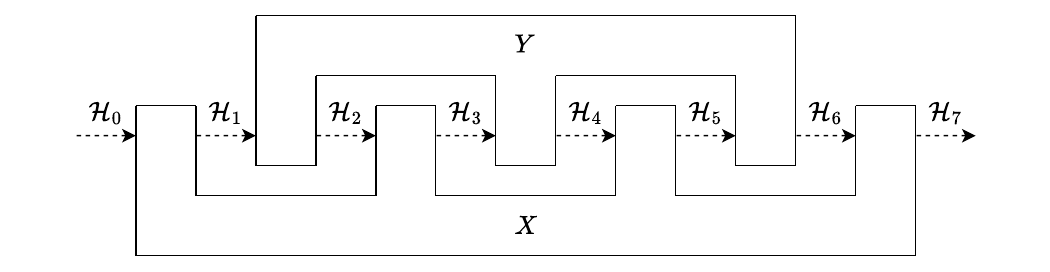}
    \caption{The combination of a $4$-comb $X$ with a $3$-comb $Y$, yielding a $1$-comb $X\star Y$ on $(\mathcal{H}_0,\mathcal{H}_7)$.}
    \label{fig-6210015}
\end{figure}

The link product provides the mathematical description of the combination of quantum combs. For instance, suppose $X$ is an $n$-comb on $(\mathcal{H}_0,\mathcal{H}_1,\ldots,\mathcal{H}_{2n-1})$ and $Y$ is an $(n-1)$-comb on $(\mathcal{H}_1,\mathcal{H}_2\,\ldots,\mathcal{H}_{2n-2})$. Then,
\begin{align}
X\star Y= \tr_{\mathcal{H}_{1:2n-2}}\!\left(X^{\textup{T}_{\mathcal{H}_{1:2n-2}}}\cdot (I_{\mathcal{H}_{2n-1}}\otimes Y \otimes I_{\mathcal{H}_0})\right)=\tr_{\mathcal{H}_{1:2n-2}}\!\left(X\cdot (I_{\mathcal{H}_{2n-1}}\otimes Y^{\textup{T}} \otimes I_{\mathcal{H}_0})\right)\nonumber
\end{align}
turns out to be a $1$-comb on $(\mathcal{H}_0,\mathcal{H}_{2n-1})$, as in the example illustrated in \cref{fig-6210015}.
The link product preserves the L\"owner order: if $X,Y\succeq 0$ then $X\star Y\succeq  0$~\cite[Theorem 2]{chiribella2009theoretical}

\paragraph{Quantum channel testers.}
A \textit{quantum channel tester} is any quantum algorithm that can make sequential, adaptive, and coherent queries to an unknown quantum channel in order to produce a classical output.
We are going to use the quantum tester formalism based on quantum comb (see, e.g., \cite{chiribella2009theoretical,bavaresco2021strict,bavaresco2022unitary}).

Consider a quantum channel tester using $n$ queries to an unknown quantum channel $\mathcal{E}$. 
Let us label the input and output systems of the $i$-th query to $\mathcal{E}$ as $\mathcal{H}_{\mathrm{A},i}$ and $\mathcal{H}_{\mathrm{B},i}$, i.e.\ the $i$-th copy of the unknown channel is a linear map from $\mathcal{L}(\mathcal{H}_{\mathrm{A},i})$ to $\mathcal{L}(\mathcal{H}_{\mathrm{B},i})$.

\begin{definition}[Channel tester]
A tester that uses $n$ queries to an unknown channel is defined as any set of linear operators $\{T_i\}_i$, with $T_i\in\mathcal{L}(\bigotimes_{j=1}^n \mathcal{H}_{\mathrm{A},j}\otimes\mathcal{H}_{\mathrm{B},j})$, such that $T_i\succeq 0$ and $\sum_i T_i$ is a quantum $(n+1)$-comb on $(\mathcal{H}_0,\mathcal{H}_{\mathrm{A},1},\mathcal{H}_{\mathrm{B},1},\ldots,\mathcal{H}_{\mathrm{A},n},\mathcal{H}_{\mathrm{B},n},\mathcal{H}_{n+1})$, where $\mathcal{H}_0\cong\mathcal{H}_{n+1}\cong\mathbb{C}$ are one-dimensional.
\end{definition}
It is known that any tester can be obtained by a coherent sequential query-access algorithm and any coherent sequential query-access algorithm can be described by a tester~\cite{chiribella2009theoretical,bavaresco2022unitary}.
When applying a tester $\{T_i\}_i$ to $n$ queries to a quantum channel $\mathcal{E}$, we get the classical outcome $i$ with probability 
\begin{equation*}
p_i=T_i\star C_\mathcal{E}^{\otimes n}= \tr(T_i (C_{\mathcal{E}}^{\otimes n})^\mathrm{T})=\tr(T_i^\mathrm{T} C_{\mathcal{E}}^{\otimes n}),
\end{equation*}
where $C_\mathcal{E}^{\otimes n}$
denotes the Choi operator of all $n$ queries to the channel $\mathcal{E}$ and $(\,\cdot\,)^{\mathrm{T}}$ represents matrix transposition.

\subsection{Schur-Weyl duality}
Let us consider a sequence of Hilbert spaces $\mathcal{H}_1,\mathcal{H}_2,\ldots,\mathcal{H}_n$ such that $\mathcal{H}_i\cong\mathbb{C}^d$ for $1\leq i\leq n$.
The Hilbert space \(\bigotimes_{i=1}^n\mathcal{H}_i\) admits representations of the symmetric group \(\mathfrak{S}_n\) (i.e.\ the group of all permutations on the set $\{1,2,\ldots,n\}$) and unitary group \(\mathbb{U}_d\) (i.e.\ the group of unitaries on $d$-dimensional Hilbert space). The unitary group acts by simultaneous ``rotation'' as \(U^{\otimes n}\) for any \(U\in \mathbb{U}_d\), while the symmetric group acts by permuting tensor factors:
\begin{equation}
\texttt{p}(\pi)\ket{\psi_1}\cdots\ket{\psi_n}=\ket{\psi_{\pi^{-1}(1)}}\cdots\ket{\psi_{\pi^{-1}(n)}},
\end{equation}
where \(\pi\in\mathfrak{S}_n\). 
The two actions \(U^{\otimes n}\) and \(\texttt{p}(\pi)\) commute with each other, hence $\bigotimes_{i=1}^n\mathcal{H}_i$ admits a representation of group \(\mathbb{U}_d\times \mathfrak{S}_n\). In particular, the Schur-Weyl duality (see, e.g., \cite{fulton2013representation}) states that
\begin{equation}\label{eq-1111224}
\bigotimes_{i=1}^n\mathcal{H}_i\stackrel{\mathfrak{S}_n\times\mathbb{U}_d}{\cong}\bigoplus_{\lambda \vdash_d\, n}\mathcal{P}_\lambda\otimes \mathcal{Q}^d_\lambda,
\end{equation}
where \(\mathcal{P}_\lambda\) and \(\mathcal{Q}^d_\lambda\) are irreducible representations of \(\mathfrak{S}_n\) and \(\mathbb{U}_d\) labeled by Young diagram $\lambda$, respectively. We denote by $\texttt{p}_\lambda(\pi)$ and $\texttt{q}_\lambda(U)$ the actions of $\pi\in\mathfrak{S}_n$ and $U\in\mathbb{U}_d$ on $\mathcal{P}_\lambda$ and $\mathcal{Q}_\lambda^d$, respectively.
In fact, $q_\lambda(\cdot)$ can be extended to be defined on all $M\in \mathcal{L}(\mathbb{C}^d)$. We may use $M_\lambda$ to denote $\texttt{q}_\lambda(M)$, and use $s_\lambda(M)$ to denote $\tr(M_\lambda)$.

Now, let us consider two sequences of Hilbert spaces $(\mathcal{H}_{\textup{A},1},\ldots,\mathcal{H}_{\textup{A},n})$ and $(\mathcal{H}_{\textup{B},1},\ldots,\mathcal{H}_{\textup{B},n})$, where $\mathcal{H}_{\textup{A},i}\cong \mathbb{C}^{d_1}$ and $\mathcal{H}_{\textup{B},j} \cong \mathbb{C}^{d_2}$. We define the action of group $\mathfrak{S}_{n}\times\mathfrak{S}_n$ on $\bigotimes_{i=1}^n \mathcal{H}_{\textup{A},i}\otimes \mathcal{H}_{\textup{B},i}$ as $\texttt{p}_\mathrm{A}(\pi_1)\otimes \texttt{p}_{\mathrm{B}}(\pi_2)$ for $(\pi_1,\pi_2)\in\mathfrak{S}_n\times\mathfrak{S}_n$, where $\texttt{p}_\mathrm{A}(\cdot)$ denotes the permutation action on $\bigotimes_{i=1}^n \mathcal{H}_{\mathrm{A},i}$, and similarly for $\texttt{p}_\mathrm{B}(\,\cdot\,)$.
We define the action of \(\mathbb{U}_{d_1}\times \mathbb{U}_{d_2}\) on \(\bigotimes_{i=1}^n \mathcal{H}_{\textup{A},i}\otimes \mathcal{H}_{\textup{B},i}\) as $(U_\mathrm{A}\otimes U_{\textup{B}})^{\otimes n}$ for $(U_{\textup{A}},U_{\textup{B}})\in \mathbb{U}_{d_1}\times\mathbb{U}_{d_2}$. 
%Note that the action of $\mathfrak{S}_n\times\mathfrak{S}_n$ commutes with the action of $\mathbb{U}_{d_1}\times\mathbb{U}_{d_2}$. 
Since the tensor product of irreducible representations of two groups $G_1,G_2$ is still irreducible with respect to $G_1\times G_2$, we can see the Schur-Weyl duality on the bipartite system:
\[\bigotimes_{i=1}^n \mathcal{H}_{\textup{A},i}\otimes \mathcal{H}_{\textup{B},i}\,\, \stackrel{\mathfrak{S}_{n}\times\mathfrak{S}_n\times\mathbb{U}_{d_1}\times\mathbb{U}_{d_2}}{\cong}\,\,\bigoplus_{\substack{\lambda\vdash_{d_1} \, n\\ \mu\vdash_{d_2} \, n}} \mathcal{P}_{\lambda}\otimes\mathcal{P}_\mu\otimes \mathcal{Q}^{d_1}_\lambda\otimes\mathcal{Q}^{d_2}_\mu,\]
where $\mathcal{P}_\lambda\otimes\mathcal{P}_\mu\otimes\mathcal{Q}^{d_1}_\lambda\otimes\mathcal{Q}^{d_2}_\mu$ is an irreducible representation of $\mathfrak{S}_n\times\mathfrak{S}_n\times \mathbb{U}_{d_1}\times\mathbb{U}_{d_2}$.

\section{Query upper bound}

\subsection{Fine-grained error bounds for isometry tomography}
Let $2\leq d<D<2d$ be two positive integers and $\alpha\coloneqq D-d$.
Let $V:\mathbb{C}^d\rightarrow\mathbb{C}^D$ be an isometry.
Suppose $\widehat{V}$ is an estimate of $V$ (which is also assumed to be an isometry).
First, we decompose the estimate $\widehat{V}$ into components inside and outside the image of $V$:
\begin{align}
\widehat{V}
&= VV^\dag \widehat{V} + \Deltaout,\nonumber 
\end{align}
where $\Deltaout\coloneqq (I-V V^\dag) \widehat{V}$ can also be viewed as the perturbation (i.e., estimation error) outside the image of $V$. We further decompose $VV^\dag \widehat{V}$ as
\begin{align}
V V^\dag \widehat{V} = zV+V \Delta_{\mathrm{in}},\nonumber
\end{align}
where $z\coloneqq \tr(V^\dag\widehat{V})/d$ and $\Deltain\coloneqq V^\dag\widehat{V}-z I$. Note that $\Deltain$ can be viewed as the perturbation (i.e., estimation error) inside the image of $V$ and satisfies $\tr(\Deltain)=0$.
Therefore, we can write
\begin{align}
\widehat{V}= zV + V\Deltain + \Deltaout.\nonumber
\end{align}
Note that $\Deltain$ and $\Deltaout$ mirrors the Type I and Type II hard instances of~\cite{chen2026quantum}, which perturb an isometry within a fixed image and outside that image, respectively.

Then, we bound the in-image error $\Deltain$ and out-of-image error $\Deltaout$ separately as follows.
\begin{lemma}\label{thm-10240443}
Let $2\leq d<D< 2d$, $\alpha=D-d$, and $n\geq 16d^2$. There exists an algorithm using $n$ parallel queries to an unknown $V\in\isochannel_{d,D}$, returning an estimate $\widehat{V}$ such that
\begin{equation}\label{eq-10240516}
\E\!\left[\|\Deltaout\|_{\mathrm{F}}^2\right] \leq \frac{2\alpha d^2}{n},\qquad \qquad \E\!\left[\|\Deltain\|_{\mathrm{F}}^2\right] \leq  (512\pi^2+6)\frac{d^5}{n^2}.
\end{equation}
\end{lemma}
Note that
\[\frac{1}{d}\left\|C_{\widehat{\mathcal{V}}}-C_{\mathcal{V}}\right\|_1=2\sqrt{1-|z|^2}=2\sqrt{(\|\Deltaout\|_{\mathrm{F}}^2+\|\Deltain\|_{\mathrm{F}}^2)/d}.\]
Thus \cref{thm-10240443} can give an $O(d^2/\varepsilon+\alpha d/\varepsilon^2)$   Choi trace norm tomography algorithm for isometry channels, recovering the result in \cite{yoshida2025quantum} (see also \cite[Lemma 8.4]{chen2026quantum}).
Furthermore, \cref{thm-10240443} gives fine-grained bounds on each $\|\Deltaout\|_\mathrm{F}$ and $\|\Deltain\|_\mathrm{F}$, which are essential for our subsequent analysis of the error after contracting a subsystem (see \cref{lemma-10242227}).

\subsection{Channel tomography in the near-boundary regime}\label{sec-10271822}
We use the following notation. Let $2\leq d_1< rd_2< 2d_1$ and $\alpha\coloneqq rd_2-d_1$. Let $\mathcal{H}_\mathrm{A}\cong\mathbb{C}^{d_1}$, $\mathcal{H}_\mathrm{B}\cong\mathbb{C}^{d_2}$, and $\mathcal{H}_\mathrm{anc}\cong\mathbb{C}^r$.
We use $\tr_\mathcal{H}\circ \mathcal{E}$ to denote the channel obtained by applying $\mathcal{E}$ followed by the tracing-out channel $\tr_\mathcal{H}$ over system $\mathcal{H}$.
Let $V:\mathcal{H}_\mathrm{A}\rightarrow\mathcal{H}_\mathrm{B}\otimes\mathcal{H}_\mathrm{anc}$ be an isometry, and $\widehat{V}$ be an estimate of $V$.

Using the ``dilation does not help'' theorem~\cite[Theorem 1.5]{chen2026quantum} and the contraction inequality:
\begin{equation}\label{eq-10240511}
\frac{1}{d_1}\left\|C_{\tr_\mathrm{anc}\circ \widehat{\mathcal{V}}}-C_{\tr_\mathrm{anc}\circ\mathcal{V}}\right\|_1 \leq \frac{1}{d_1}\left\|C_{\widehat{\mathcal{V}}}-C_{{\mathcal{V}}}\right\|_1,
\end{equation}
we can recover the query upper bound $O(d_1^2/\varepsilon+(rd_2-d_1)d_1/\varepsilon^2)$ for quantum channel tomography~\cite[Theorem 3.7]{chen2026quantum}. 
However, \cref{eq-10240511} is still too coarse to bound the error $\frac{1}{d_1}\|C_{\tr_\mathrm{anc}\circ \widehat{\mathcal{V}}}-C_{\tr_\mathrm{anc}\circ\mathcal{V}}\|_1$. 
For this, we show that \cref{eq-10240516} can be used to give a finer analysis for the error after tracing out the ancilla system.

\begin{lemma}\label{lemma-10242227}
Let $\alpha=rd_2-d_1$ and $\beta= \min\{r,\alpha d_2\}$. Then,
\begin{align}
    \frac{1}{d_1}\left\|C_{\tr_\mathrm{anc}\circ\widehat{\mathcal{V}}}-C_{\tr_\mathrm{anc}\circ\mathcal{V}}\right\|_1\leq \frac{2}{\sqrt{d_1}}\|\Deltain\|_\mathrm{F}+\frac{2\sqrt{d_2 \beta}}{d_1}\|\Deltaout\|_\mathrm{F}+\frac{2}{d_1}\left(\|\Deltain\|_\mathrm{F}^2+\|\Deltaout\|_\mathrm{F}^2\right).\label{eq-10241742}
\end{align}
\end{lemma}
\begin{proof}
Note that 
\begin{align}
C_{\tr_\mathrm{anc}\circ \widehat{\mathcal{V}}}&= \tr_{\mathrm{anc}}\kettbbra{\widehat{V}}{\widehat{V}}\nonumber \\
&=|z|^2\tr_\mathrm{anc}\kettbbra{V}{V}+\tr_\mathrm{anc}\kettbbra{\Delta}{\Delta}+z\tr_\mathrm{anc}\kettbbra{V}{\Delta}+z^*\tr_\mathrm{anc}\kettbbra{\Delta}{V},\nonumber
\end{align}
where we define $\Delta\coloneqq V\Deltain+\Deltaout$.

We first bound $\|\tr_\mathrm{anc}\kettbbra{V}{\Deltaout}\|_1$. 
For $X:\mathcal{H}_\mathrm{A}\rightarrow \mathcal{H}_\mathrm{B}\otimes\mathcal{H}_\mathrm{anc}$, we can write $X=\sum_{i=1}^rX_i\otimes \ket{i}_\mathrm{anc}$, and define $A_X$ as the matrix 
\[A_X\coloneqq \sum_{i=1}^r\kett{X_i}\bra{i}_\mathrm{anc}:\mathcal{H}_\mathrm{anc}\rightarrow\mathcal{H}_\mathrm{A}\otimes \mathcal{H}_\mathrm{B}.\]
Then we can see that 
\[\tr_\mathrm{anc}\kettbbra{V}{\Deltaout}=A_V A_{\Deltaout}^\dag.\]
Since $A_V^\dag A_V=(\tr_\mathrm{B}VV^\dag)^\mathrm{T}$ and $VV^\dag\preceq I_{rd_2}$, we know that $\|A_V\|_{\mathrm{op}}\leq \sqrt{d_2}$. Also, $A^\dag_{\Deltaout} A_\Deltaout=(\tr_\mathrm{B} \Deltaout\Delta^\dag_{\mathrm{out}})^\mathrm{T}$ and note that the image of $\Deltaout$ lies in a subspace of dimension $\alpha$, thus the rank of $\tr_\mathrm{B} \Deltaout\Delta^\dag_{\mathrm{out}}$ is at most $\alpha d_2$, which means $\mathrm{rank} A_\Deltaout\leq \alpha d_2$. Obviously, we also have $\mathrm{rank} A_{\Deltaout}\leq r$, thus $\mathrm{rank}A_{\Deltaout}\leq \beta$. 
Therefore,
%Using the reshaping notation and identities from \cref{eq:Kraus,eq:reshaping-identities}, 
%we have
%$$\tr_{\mathrm{anc}}\kettbbra{V}{\Deltaout}=A_V A_{\Deltaout}^\dag.$$ By \cref{eq:Kraus-Choi}, $\|A_V\|_{\mathrm{op}}\leq\sqrt{d_2}$. Since $\Deltaout$ has image dimension at most $\alpha$, \cref{eq:reshaping-rank} gives $\rank A_{\Deltaout}\leq\beta$. Together with the Frobenius-norm identity in \cref{eq:reshaping-inner-product}, these bounds give
\begin{align}
    \|A_V A_{\Deltaout}^\dagger\|_1\leq \sqrt{d_2}\|A_\Deltaout\|_1\leq \sqrt{d_2\beta}\|A_\Deltaout\|_\mathrm{F}=\sqrt{d_2\beta}\|\Deltaout\|_\mathrm{F}.\label{eq-10241743}
\end{align}
Then we bound $\|\tr_\mathrm{anc}\kettbbra{V}{V\Deltain}\|_1$:
\begin{align}
    \|\tr_\mathrm{anc}\kettbbra{V}{V\Deltain}\|_1\leq \|\kettbbra{V}{V\Deltain}\|_1 = \sqrt{d_1}\|\Deltain\|_\mathrm{F}.\label{eq-10241744}
\end{align} 
Also, note that 
\begin{align}
\|\tr_\mathrm{anc}\kettbbra{\Delta}{\Delta}\|_1\leq \|\kettbbra{\Delta}{\Delta}\|_1=\|\Delta\|^2_\mathrm{F}=d_1(1-|z|^2).\label{eq-10241745}
\end{align}
Therefore, combining \cref{eq-10241743,eq-10241744,eq-10241745}, we have
\begin{align}
\frac{1}{d_1}\left\|C_{\tr_\mathrm{anc}\circ \widehat{\mathcal{V}}}-C_{\tr_\mathrm{anc}\circ\mathcal{V}}\right\|_1&\leq 2(1-|z|^2)+2|z|\frac{\sqrt{d_2\beta}}{d_1}\|\Deltaout\|_\mathrm{F}+2|z|\frac{1}{\sqrt{d_1}}\|\Deltain\|_\mathrm{F}.\nonumber
\end{align}
Then using $1-|z|^2=\frac{1}{d_1}(\|\Deltain\|_\mathrm{F}^2+\|\Deltaout\|_\mathrm{F}^2)$ and $|z|\leq 1$, we can conclude \cref{eq-10241742}.
\end{proof}

Then, we can prove the query upper bound.
\begin{theorem}\label{thm-10300424}
Suppose $d_1<rd_2<2d_1$ and let $\alpha=rd_2-d_1<d_1$.
There is an algorithm that uses
\[O\!\left(\frac{d_1^2}{\varepsilon}+ \frac{\min\{\alpha d_1,\alpha^2 d_2^2\}}{\varepsilon^2}\right)\]
parallel queries to $\mathcal{E}\in\qchannel_{d_1,d_2}^r$ and outputs an estimate $\widehat{\mathcal{E}}$ such that $\frac{1}{d_1}\|C_{\widehat{\mathcal{E}}}-C_{\mathcal{E}}\|_1\leq \varepsilon$ with probability at least $2/3$.
\end{theorem}
\begin{proof}
Combining \cref{thm-10240443} with \cref{lemma-10242227}, we know that there is an algorithm that uses $O(\frac{d_1^2}{\varepsilon}+ \frac{\min\{\alpha d_1,\alpha^2 d_2^2\}}{\varepsilon^2})$ parallel queries to $V\in \isochannel_{d_1,rd_2}$ and outputs an estimate $\widehat{V}$ such that $\frac{1}{d_1}\|C_{\tr_\mathrm{anc}\circ\widehat{\mathcal{V}}}-C_{\tr_\mathrm{anc}\circ\mathcal{V}}\|_1\leq O(\varepsilon)$. Here, we treat $V$ as a Stinespring dilation of $\tr_\mathrm{anc}\circ\mathcal{V}\in\qchannel_{d_1,d_2}^r$, and $\tr_\mathrm{anc}\circ\widehat{\mathcal{V}}$ as an estimate of $\tr_\mathrm{anc}\circ\mathcal{V}$. Thus, by \cite[Theorem 1.5]{chen2026quantum}, there is also an algorithm that uses the same number of queries to $\mathcal{E}\in\qchannel_{d_1,d_2}^r$ and outputs an estimate to within   Choi trace norm error $O(\varepsilon)$, with probability at least $2/3$.
\end{proof}

\subsection{Proof of \cref{thm-10240443}}
In this proof, we build upon the isometry estimation framework of \cite{yoshida2025quantum} (see also  \cite{yang2020optimal} for the unitary estimation) with different error analysis. Rather than optimizing the overall estimation fidelity, we establish separate bounds for the components of the estimation error inside and outside the image of the unknown isometry.

\paragraph{The algorithm.}
Consider the Schur-Weyl decomposition $(\mathbb{C}^k)^{\otimes n}\cong \bigoplus_{\lambda\vdash_k n} \mathcal{P}_{\lambda}\otimes\mathcal{Q}_\lambda^k$.
Then, the algorithm
\begin{enumerate}
\item Prepare a probe state in the Schur-Weyl basis of $(\mathbb{C}^d)^{\otimes n}\otimes(\mathbb{C}^d)^{\otimes n}$
\[\bigoplus_{\lambda\vdash_d n} \frac{v_\lambda}{\sqrt{\dim \mathcal{Q}_\lambda^d}} \ket{\bot_\lambda}\otimes \kett{I_{\mathcal{Q}_\lambda^d}},\]
where $v_\lambda$ are real numbers such that $\sum_{\lambda\vdash_d n}v_\lambda^2=1$ and $\ket{\bot_\lambda}\in\mathcal{P}_{\lambda}\otimes\mathcal{P}_\lambda$ is an arbitrary unit vector.

\item Applies $n$ parallel queries of the unknown isometry $V^{\otimes n}$ to the first part of the probe state, resulting
\begin{align}
\bigoplus_{\lambda\vdash_d n}\frac{v_\lambda}{\sqrt{\dim \mathcal{Q}_\lambda^d}}\ket{\bot_\lambda}\otimes \kett{V_\lambda},\nonumber
\end{align}
where $V_\lambda:\mathcal{Q}_\lambda^d\rightarrow\mathcal{Q}_\lambda^D$ is an isometry (this can be seen by considering the action of $V^{\otimes n}$ on the Schur-Weyl basis and by noting that $V^{\otimes n}$ commutes with the permutation).

\item Perform the POVM $\{\ketbra{\zeta_{\widehat{V}}}{\zeta_{\widehat{V}}}\}_{\widehat{V}\in\isochannel_{d,D}}$, where
\[\ket{\zeta_{\widehat{V}}}\coloneqq \bigoplus_{\lambda\vdash_d n}\sqrt{\dim \mathcal{Q}_\lambda^D}\ket{\bot_\lambda}\otimes\kett{\widehat{V}_\lambda}.\]
This is a POVM since 
\[\E_{\widehat{V}\sim\isochannel_{d,D}}\ketbra{\zeta_{\widehat{V}}}{\zeta_{\widehat{V}}}=\bigoplus_{\lambda\vdash_d n}\ketbra{\bot_\lambda}{\bot_\lambda}\otimes I_{\mathcal{Q}_\lambda^D}\otimes I_{\mathcal{Q}_\lambda^d}\preceq I,\]
where we use that $U\widehat{V}$ and $\widehat{V}$ have the same distribution for $U\in\mathbb{U}_{D}$ and $\widehat{V}\sim \isochannel_{d,D}$, with Schur's lemma. 
We note that this POVM can also be interpreted as a Pretty Good Measurement (see, e.g., \cite{chen2025inverse}).
\end{enumerate}
Therefore, on input $V$, the algorithm outputs an estimate $\widehat{V}$ with probability density
\begin{align}
\Pr[\widehat{V}|V]=\left|\sum_{\lambda\vdash_d n}v_\lambda\sqrt{\frac{\dim\mathcal{Q}_\lambda^D}{\dim \mathcal{Q}_\lambda^d}} s_\lambda(V^\dag\widehat{V})\right|^2,\label{eq-10280133}
\end{align}
where $s_\lambda(V^\dag\widehat{V})=\tr(V_\lambda^\dag \widehat{V}_\lambda)$.

\paragraph{Expressions for the errors.}
Set $M\coloneqq V^\dag\widehat{V}$. Note that
\begin{align}
\|\Deltaout\|_\mathrm{F}^2=d-\tr(M^\dag M), \qquad\qquad \|\Deltain\|_\mathrm{F}^2=\tr(M^\dag M)-\frac{|\tr M|^2}{d}.\label{eq-10280508}
\end{align}

First, we compute $\E \tr(M^\dag M)$ w.r.t. fixed $V$ and the randomness of $\widehat{V}$ according to \cref{eq-10280133} as follows
\begin{align}
\E \tr(M^\dag M)&= \E_{\widehat{V}\sim\isochannel_{d,D}} \left|\sum_{\lambda\vdash_d n}v_\lambda\sqrt{\frac{\dim\mathcal{Q}_\lambda^D}{\dim \mathcal{Q}_\lambda^d}} s_\lambda(M)\right|^2 \tr(M^\dag M)\nonumber \\
&= \sum_{\lambda\vdash_d n}v_\lambda^2\frac{\dim\mathcal{Q}_\lambda^D}{(\dim\mathcal{Q}_\lambda^d)^2}\E_{\widehat{V}\sim\isochannel_{d,D}} s_\lambda(M^\dag M)\tr(M^\dag M)\label{eq-10280329}\\
&=\sum_{\lambda\vdash_d n}v_\lambda^2\frac{\dim\mathcal{Q}_\lambda^D}{(\dim\mathcal{Q}_\lambda^d)^2} \sum_{\mu=\lambda+\Box} \E_{\widehat{V}\sim\isochannel_{d,D}}s_\mu(M^\dag M)\label{eq-10280354}\\
&=\sum_{\lambda\vdash_d n}v_\lambda^2\frac{\dim\mathcal{Q}_\lambda^D}{(\dim\mathcal{Q}_\lambda^d)^2} \sum_{\mu=\lambda+\Box} \frac{(\dim\mathcal{Q}_\mu^d)^2}{\dim\mathcal{Q}_\mu^D}.\label{eq-10280355}
\end{align}
where \cref{eq-10280329} first uses the fact that $M$ and $MU$ has the same distribution for any $U\in\mathbb{U}_d$ and $\widehat{V}\sim\isochannel_{d,D}$, and then uses \cref{eq-10280330}, \cref{eq-10280354} is by Pieri's rule where $\mu=\lambda+\Box$ means $\mu$ is a valid Young diagram obtained by adding one box to $\lambda$, and \cref{eq-10280355} is due to \cref{eq-10280356}.
For convenience, we define
\begin{align}
u_{\lambda,i}\coloneqq \frac{1}{\lambda_i-i+D+1},\qquad f_{\lambda,i}=\sqrt{\frac{\lambda_i-i+d+1}{\lambda_i-i+D+1}}=\sqrt{1-\alpha u_{\lambda,i}}.\label{eq-10290017}
\end{align}
When $\lambda+\bm{e}_i$ is a valid partition (i.e., adding a box to the $i$-th row of $\lambda$), set 
\[c_{\lambda,i}\coloneqq \frac{\dim \mathcal{Q}_{\lambda+\bm{e}_i}^d}{\dim \mathcal{Q}_\lambda^d},\]
and otherwise set $c_{\lambda,i}=0$. The hook-content dimension formula and the Pieri's rule give, respectively,
\begin{align}
\frac{\dim \mathcal{Q}_{\lambda+\bm{e}_i}^D}{\dim\mathcal{Q}_\lambda^D}=\frac{c_{\lambda,i}}{f_{\lambda,i}^2},\qquad \sum_{i=1}^d c_{\lambda,i}=d. \label{eq-10280452}
\end{align}
Thus each summand of \cref{eq-10280355} is simply $c_{\lambda,i}f_{\lambda,i}^2$, and we have
\begin{align}
\E \tr(M^\dag M)=\sum_{\lambda\vdash_d n}v_\lambda^2\sum_{i=1}^d c_{\lambda,i}f_{\lambda,i}^2=d-\alpha\sum_{\lambda\vdash_d n}v_\lambda^2\sum_{i=1}^d c_{\lambda,i} u_{\lambda,i}.\label{eq-10280510} 
\end{align}

Then, we compute $\E |\tr M|^2$ w.r.t. fixed $V$ and the randomness of $\widehat{V}$ according to \cref{eq-10280133} as follows
\begin{align}
\E |\tr M|^2&=\E_{\widehat{V}\sim\isochannel_{d,D}}\left|\sum_{\lambda\vdash_d n}v_\lambda\sqrt{\frac{\dim\mathcal{Q}_\lambda^D}{\dim \mathcal{Q}_\lambda^d}}s_\lambda(M)\tr M\right|^2\nonumber \\
&=\E_{\widehat{V}\sim\isochannel_{d,D}}\left|\sum_{\mu\vdash_d n+1}\left(\sum_{i:\mu-\bm{e}_i \textup{ valid}}v_{\mu-\bm{e}_i}\sqrt{\frac{\dim\mathcal{Q}_{\mu-\bm{e}_i}^D}{\dim\mathcal{Q}_{\mu-\bm{e}_i}^d}}\right)s_{\mu}(M)\right|^2\label{eq-10280440}\\
&=\sum_{\mu\vdash_d n+1}\frac{\dim\mathcal{Q}_\mu^d}{\dim\mathcal{Q}_\mu^D}\left(\sum_{i:\mu-\bm{e}_i \textup{ valid}}v_{\mu-\bm{e}_i}\sqrt{\frac{\dim\mathcal{Q}_{\mu-\bm{e}_i}^D}{\dim\mathcal{Q}_{\mu-\bm{e}_i}^d}}\right)^2,\label{eq-10280450}
\end{align}
where \cref{eq-10280440} is due to Pieri's rule and \cref{eq-10280450} is due to \cref{eq-10280451}.
Finally, using \cref{eq-10280452}, we know that 
\begin{align}
\E|\tr M|^2=\sum_{\mu\vdash_d n+1}\left(\sum_{i=1}^d v_{\mu-\bm{e}_i}f_{\mu-\bm{e}_i,i}\right)^2,\label{eq-10280509}
\end{align}
where we adopt the convention that an invalid subtraction contributes zero.
Note that $\E \|\Deltaout\|_\mathrm{F}^2$ and $\E \|\Deltain\|_\mathrm{F}^2$ can be computed using \cref{eq-10280508} with \cref{eq-10280510} and \cref{eq-10280509}.

\paragraph{Define the amplitudes $v_\lambda$.}
Next, we choose good $v_\lambda$ for \cref{eq-10280510} and \cref{eq-10280509}, using a similar strategy as that in \cite{yoshida2025quantum} (see also \cite{yang2020optimal}).
Set 
\begin{align}
g\coloneqq \left\lfloor\frac{n}{8d^2}\right\rfloor, \qquad n-\frac{d(d-1)}{2}g=dh+s,\quad\textup{where}\,\, 0\leq s<d.\nonumber
\end{align}
As $n\geq 16d^2$, we have $g\geq 2$ and $g\geq n/(16d^2)$. For $x=(x_1,\ldots,x_{d-1})\in \{0,\ldots,g-1\}^{d-1}$, put
\begin{equation}\label{eq-10290255}
\begin{split}
    \lambda_i(x)&=h+g(d-i)+\mathbbm{1}_{i\leq s}+x_i,\qquad 1\leq i<d,\\
    \lambda_d(x)&=h-\sum_{j=1}^{d-1}x_j.
\end{split}
\end{equation}
Every $\lambda(x)$ has $n$ boxes. For $i<d-1$, the adjacent row gap $\lambda_i(x)-\lambda_{i+1}(x)$ is at least $g-(g-1)=1$, and the last gap is at least $g$. Thus all the $\lambda(x)$ are strictly decreasing partitions, and all $d$ additions $\lambda(x)+\bm{e}_i$ are valid. Moreover,
\begin{align}
\lambda_d(x)&\geq \frac{n}{d}-\frac{3(d-1)}{2}g+(d-1)-\frac{s}{d}\geq\frac{13n}{16d}\geq \frac{3n}{4d}.\label{eq-10290345}
\end{align}
In particular, on the box and on its one-box additions, we have 
\begin{align}
0\leq u_{\lambda,i}\leq \frac{2d}{n}.\label{eq-10290019}
\end{align}

Now we define
\begin{equation}
a_t\coloneqq 
\begin{cases}
\sqrt{\frac{2}{g}}\sin\!\left(\frac{\pi(2t+1)}{2g}\right), \qquad &0\leq t<g,\\
0,&\textup{otherwise}.
\end{cases}
\end{equation}
Set the amplitudes to be
\begin{align}
v_{\lambda(x)}\coloneqq \prod_{j=1}^{d-1}a_{x_j}, \qquad \textup{and} \qquad v_\lambda\coloneqq 0\quad \textup{for other }\lambda\vdash_d n.\label{eq-10290418}
\end{align}
The identity $\sum_{t=0}^{g-1}a_t^2=1$ proves $\sum_{\lambda\vdash_d n} v_\lambda^2=1$. The nearest-neighbor overlap is exactly
\begin{align}
\sum_{t=0}^{g-2} a_t a_{t+1}&=\frac{1}{g}+\frac{g-1}{g}\cos\frac{\pi}{g}=1-\varepsilon_g,\label{eq-10290419}
\end{align}
where $\varepsilon_g\coloneqq \frac{g-1}{g}(1-\cos\frac{\pi}{g})\leq \frac{\pi^2}{2g^2}$.

\paragraph{Bounding the outer error.}
Therefore, by \cref{eq-10280508} and \cref{eq-10280510}, we know that
\begin{align}
\E \|\Deltaout\|_\mathrm{F}^2&=\alpha\sum_{\lambda\vdash_d n}v_\lambda^2\sum_{i=1}^d c_{\lambda,i}u_{\lambda,i}\nonumber\\
&\leq \alpha\sum_{\lambda\vdash_d n}v_\lambda^2 d \frac{2d}{n}=\frac{2\alpha d^2}{n},
\end{align}
where we used $u_{\lambda,i}\leq 2d/n$ (see \cref{eq-10290019}) and $\sum_{i=1}^dc_{\lambda,i}=d$ (see \cref{eq-10280452}).

\paragraph{Bounding the inner error.}
Set
\[R_0\coloneqq \sum_{\lambda\vdash_d n} v_\lambda^2\sum_{i=1}^d f_{\lambda,i}^2=d-\alpha\sum_{\lambda\vdash_d n} v_\lambda^2\sum_{i=1}^d u_{\lambda,i},\]
where we used \cref{eq-10290017}.
By \cref{eq-10280508}, we have
\begin{align}
\E\|\Deltain\|_\mathrm{F}^2&=\E\tr(M^\dag M)-\frac{1}{d}\E|\tr M|^2 \nonumber \\
&=\left(\E\tr(M^\dag M)-R_0\right) + \left(R_0 - \frac{1}{d}\E|\tr M|^2\right).\label{eq-10290453}
\end{align}
We will bound these two terms separately.

\textbf{First term}: $\E\tr(M^\dag M)-R_0$. 
We define
\[\ell_i\coloneqq \lambda_i+d-i.\]
Using Weyl's dimension formula, we know that $c_{\lambda,i}=\prod_{j\neq i}(\ell_i-\ell_j+1)/(\ell_i-\ell_j)$.
These numbers are the residues of the rational function $F(z)=\prod_{j}(z-\ell_j+1)/(z-\ell_j)$ at $z=\ell_i$. Since $\ell_i$ are pairwise different and $\lim_{z\rightarrow\infty}F(z)=1$, we know its partial-fraction expansion must be $F(z)=1+\sum_{i=1}^d\frac{c_{\lambda,i}}{z-\ell_i}$. Evaluating at $z=-\alpha-1$ therefore yields
\[\sum_{i=1}^d c_{\lambda,i} u_{\lambda,i}=1-\prod_{i=1}^d (1-u_{\lambda,i}).\]
Combining this with \cref{eq-10280510}, we obtain
\[\E\tr(M^\dag M)-R_0=\alpha \sum_{\lambda\vdash_d n}v_\lambda^2 \left[\sum_{i=1}^d u_{\lambda,i}-1+\prod_{i=1}^d (1-u_{\lambda,i})\right].\]
Using the inequality
\[0\leq \sum_{i} u_{i}-1+\prod_i (1-u_{i})\leq \sum_{i<j}u_iu_j\leq \frac{1}{2}\left(\sum_i u_i\right)^2,\qquad (0\leq u_i\leq 1)\]
with \cref{eq-10290019}, we obtain 
\begin{align}
0\leq \E\tr(M^\dag M)-R_0\leq \frac{\alpha}{2}\left(\frac{2d^2}{n}\right)^2=\frac{2\alpha d^4}{n^2}.\label{eq-10290454}
\end{align}

\textbf{Second term}: $R_0 - \frac{1}{d}\E|\tr M|^2$. 
%For each $\mu\vdash_d n+1$, set
%\[w_{\mu,i}\coloneqq v_{\mu-\bm{e}_i}f_{\mu-\bm{e}_i,i},\]
%where for invalid $\mu-\bm{e}_i$, it contributes zero. 
Note that $\sum_{\mu,i}v_{\mu-\bm{e}_i}^2f_{\mu-\bm{e}_i,i}^2=R_0$, since every $\lambda$ in \cref{eq-10290255} admits adding a box for all $d$ rows (for invalid $\mu-\bm{e}_i$, it contributes zero). Combining this with \cref{eq-10280509} and the identity $\sum_{1\leq i<j\leq d}(x_i-x_j)^2=d\sum_{i=1}^d x_i^2-(\sum_{i=1}^d x_i)^2$, we get
\begin{align}
R_0 - \frac{1}{d}\E|\tr M|^2=\frac{1}{d}\sum_{\mu\vdash_d n+1}\sum_{1\leq i<j\leq d}(v_{\mu-\bm{e}_i}f_{\mu-\bm{e}_i,i}-v_{\mu-\bm{e}_j}f_{\mu-\bm{e}_j,j})^2.\label{eq-10290359}
\end{align}
%Note that $\mu_i+D-i\geq \mu_d\geq 3n/(4d)$ by \cref{eq-10290345}. 
Thus, we only need to consider $\mu\vdash_d n+1$ such that $v_{\mu-\bm{e}_k}\neq 0$ for some $k\in[d]$.
Using \cref{eq-10290019}, for all $i,j$,
\begin{align}
1-\frac{2\alpha d}{n}\leq f_{\mu-\bm{e}_i,i}\leq 1, \qquad |f_{\mu-\bm{e}_i,i}-f_{\mu-\bm{e}_j,j}|\leq \frac{2\alpha d}{n}.\label{eq-10290405} 
\end{align}
Consequently, using \cref{eq-10290405} on \cref{eq-10290359}, we have
\begin{align}
R_0-\frac{1}{d}\E|\tr M|^2 &= \frac{1}{d} \sum_{\mu\vdash_d n+1}\sum_{1\leq i<j\leq d}\left[(v_{\mu-\bm{e}_i}-v_{\mu-\bm{e}_j})f_{\mu-\bm{e}_i,i}+v_{\mu-\bm{e}_j}(f_{\mu-\bm{e}_i,i}-f_{\mu-\bm{e}_j,j})\right]^2 \nonumber \\
&\leq \frac{2}{d} \sum_{\mu\vdash_d n+1}\sum_{1\leq i<j\leq d}(v_{\mu-\bm{e}_i}-v_{\mu-\bm{e}_j})^2+\frac{2}{d}\left(\frac{2\alpha d}{n}\right)^2\sum_{\mu\vdash_d n+1}\sum_{1\leq i<j\leq d}v_{\mu-\bm{e}_j}^2\nonumber \\
&\leq \frac{2}{d}\sum_{\mu\vdash_d n+1}\sum_{1\leq i<j\leq d}(v_{\mu-\bm{e}_i}-v_{\mu-\bm{e}_j})^2+\frac{4\alpha^2d^3}{n^2},\label{eq-10290449}
\end{align}
where in the last step we used $\sum_{\mu} v^2_{\mu-\bm{e}_j}=1$ for each $j$ since every supported partition admits all $d$ additions.
Using \cref{eq-10290418} and \cref{eq-10290419}, we obtain
\[\sum_{\mu\vdash_d n+1} v_{\mu-\bm{e}_i}v_{\mu-\bm{e}_j}=\begin{cases}
1-\varepsilon_g,\qquad &i<j=d,\\
(1-\varepsilon_g)^2, & i<j<d.
\end{cases}\]
Indeed, two parent boxes differ in one sine-coordinate in the first case and in two in the second. As $\sum_{\mu\vdash_d n+1} v_{\mu-\bm{e}_i}^2=1$, this proves
\begin{align}
    \sum_{\mu\vdash_d n+1}\sum_{1\leq i<j\leq d} (v_{\mu-\bm{e}_i}-v_{\mu-\bm{e}_j})^2&=2(d-1)\varepsilon_g+(d-1)(d-2)(1-(1-\varepsilon_g)^2)\nonumber\\
    &\leq 2d^2\varepsilon_g\leq \frac{d^2\pi^2}{g^2}.\nonumber
\end{align}
Combining this with \cref{eq-10290449} gives
\[R_0-\frac{1}{d}\E|\tr M|^2\leq \frac{2d\pi^2}{g^2}+\frac{4\alpha^2d^3}{n^2}\leq \frac{512\pi^2 d^5}{n^2}+\frac{4\alpha^2d^3}{n^2},\]
since $g\geq n/(16d^2)$. Further combining this with \cref{eq-10290453} and \cref{eq-10290454}, we obtain
\[\E\|\Deltain\|_\mathrm{F}^2\leq \frac{512\pi^2 d^5}{n^2}+\frac{4\alpha^2d^3}{n^2}+\frac{2\alpha d^4}{n^2}\leq (512\pi^2+6)\frac{d^5}{n^2},\]
where we used $\alpha<d$.

\subsubsection{Technical lemma}
\begin{lemma}
Let $M=V^\dag \widehat{V}$, where $V$ is fixed. Then,
\begin{align}
\E_{\widehat{V}\sim\isochannel_{d,D}} s_\lambda(M)s_\mu^*(M) &= \mathbbm{1}_{\lambda=\mu}\frac{\dim\mathcal{Q}_\lambda^d}{\dim\mathcal{Q}_\lambda^D},\label{eq-10280451}\\
\E_{U\sim\mathbb{U}_d}s_\lambda(MU)s_\mu^*(MU) &= \mathbbm{1}_{\lambda=\mu}\frac{s_\lambda(M^\dag M)}{\dim\mathcal{Q}_\lambda^d},\label{eq-10280330}\\
\E_{\widehat{V}\sim\isochannel_{d,D}}s_\lambda(M^\dag M) &=\frac{(\dim \mathcal{Q}_\lambda^d)^2}{\dim\mathcal{Q}_\lambda^D}.\label{eq-10280356}
\end{align}
\end{lemma}
\begin{proof}
Let $\texttt{q}_\lambda(U)$ denote the action of
$U\in \mathbb{U}_k$ on $\mathcal{Q}_\lambda^k$. Since $\widehat V$ is Haar-random,
we may write $\widehat V=WV$, where $W$ is Haar-random in $\mathbb{U}_D$.
Let $V_\lambda:\mathcal{Q}_\lambda^d\to \mathcal{Q}_\lambda^D$ be the isometry
induced by $V$. Then
\[
s_\lambda(M)
=\operatorname{tr}\bigl(
V_\lambda^\dagger \texttt{q}_\lambda(W)V_\lambda
\bigr)=\tr(V_\lambda V_\lambda^\dag \texttt{q}_\lambda(W)).
\]
By Schur orthogonality,
\begin{align}
\E_{\widehat V\sim\isochannel_{d,D}}
s_\lambda(M)s_\mu^*(M)
&=\E_{W\sim\mathbb{U}_D}\tr(V_\lambda V_\lambda^\dag \texttt{q}_\lambda(W))\tr(V_\mu V_\mu^\dag \texttt{q}_\mu(W))^* \nonumber \\
&=\mathbbm{1}_{\lambda=\mu}
\frac{\operatorname{tr}(V_\lambda V_\lambda^\dagger V_\lambda V_\lambda^\dagger)}
{\dim \mathcal{Q}_\lambda^D}=\mathbbm{1}_{\lambda=\mu}
\frac{\dim \mathcal{Q}_\lambda^d}{\dim \mathcal{Q}_\lambda^D}.\label{eq-10291658}
\end{align}
Similarly, for fixed $M$, Schur orthogonality over $\mathbb{U}_d$ gives
\begin{align}
\E_{U\sim\mathbb{U}_d}
s_\lambda(MU)s_\mu^*(MU)
&=\mathbbm{1}_{\lambda=\mu}
\frac{s_\lambda(M^\dagger M)}{\dim \mathcal{Q}_\lambda^d}.\label{eq-10291657}
\end{align}
Finally, since $\widehat VU$ has the same distribution as
$\widehat V$ for any $U\in\mathbb{U}_d$, averaging the \cref{eq-10291657} over $\widehat V$
and comparing with \cref{eq-10291658} yields
\[
\E_{\widehat{V}\sim\isochannel_{d,D}}s_\lambda(M^\dagger M)
=\frac{(\dim \mathcal{Q}_\lambda^d)^2}{\dim \mathcal{Q}_\lambda^D}.
\]
\end{proof}

\section{Hardness of isometry channel discrimination}\label{sec:hard-iso}
In this section, we introduce a discrimination task for isometries in $\isochannel_{d,D}$ where $d,D$ are positive integers such that $d<D<2d$. Let $\alpha\coloneqq D-d$.

Suppose $d'\leq d$ is a positive integer, we consider the decomposition of the input and output spaces
\begin{align}
\mathbb{C}^{d}= \mathbb{C}^{d-d'}\oplus \mathbb{C}^{d'}, \qquad \mathbb{C}^{D}=\mathbb{C}^{d-d'}\oplus \mathbb{C}^{d'+\alpha}.\label{eq-10270000}
\end{align}
For any $P\in\Proj_{d'}(d'+\alpha)$ (i.e., $P$ is a rank-$d'$ orthogonal projector in the background space $\mathbb{C}^{d'+\alpha}$), we fix an isometry $W_P:\mathbb{C}^{d'}\rightarrow \mathbb{C}^{d'+\alpha}$ with $W_PW_P^\dag = P$. For $U\in\mathbb{U}_{d'}$, we define
\[V_{P,U}\coloneqq I_{d-d'}\oplus W_P U:\mathbb{C}^d\rightarrow\mathbb{C}^{D}.\]
We can  easily verify that $V_{P,U}$ is an isometry.

Then, we consider the following isometry discrimination task with a Haar-random mask.
\begin{problem}[Isometry discrimination with a Haar-random mask]\label{prob-10260459}
Suppose $\mathcal{N}\subseteq \Proj_{d'}(d'+\alpha)$ is a given finite set of orthogonal projectors.
The referee draws a projector $P\in\mathcal{N}$ uniformly, and also independently draws a Haar-random $U\sim\mathbb{U}_{d'}$ and provides you queries to the isometry $V_{P,U}$. The task for you is to identify $P$.
\end{problem}

The following theorem provides the hardness of solving \cref{prob-10260459}. 
\begin{theorem}[Hardness of discrimination]\label{thm:hard-isom-disc}
The success probability of any $n$-query algorithm for \cref{prob-10260459} is bounded as
\[\Pr[\textup{success}]\leq \frac{1}{|\mathcal{N}|}\left(e+\frac{2en}{d'(d'+\alpha)}\right)^{\alpha d'}.\]
\end{theorem}
\begin{proof}
Suppose the algorithm is represented by the tester $\{T_P\}_{P\in\mathcal{N}}$.
Then, the success probability can be expressed as
\begin{align}
\Pr[\textup{success}]=\frac{1}{|\mathcal{N}|}\sum_{P\in\mathcal{N}}T_P\star \E_{U\sim\mathbb{U}_{d'}}C_{\mathcal{V}_{P,U}}^{\otimes n}.\nonumber 
\end{align}
For any $P\in \Proj_{d'}(d'+\alpha)$, we define 
\begin{align}
\Gamma_P\coloneqq \E_{U\sim\mathbb{U}_{d'}}C_{\mathcal{V}_{P,U}}^{\otimes n}.\label{eq-10260541}
\end{align}
We also define
\begin{align}
\Gamma\coloneqq \E_{P\sim \Proj_{d'}(d'+\alpha)} \Gamma_P.\label{eq-10260542}
\end{align}
Note that since $C_{\mathcal{V}_{P,U}}^{\otimes n}$ is a quantum comb, so $\Gamma$ is a convex combination of quantum combs, which is also a quantum comb.
Furthermore, \cref{lemma-10260526} shows that $\Gamma$ can serve as a good universal (i.e., $P$-independent) upper bound for all $\Gamma_P$:
\begin{align}
\Gamma_P\preceq \left(e+\frac{2en}{d'(d'+\alpha)}\right)^{\alpha d'}\Gamma.\nonumber
\end{align}
Therefore, we can bound the success probability as
\begin{align}
\Pr[\textup{success}]&=\frac{1}{|\mathcal{N}|}\sum_{P\in\mathcal{N}}T_P\star \Gamma_P \nonumber \\
&\leq \frac{1}{|\mathcal{N}|}\left(e+\frac{2en}{d'(d'+\alpha)}\right)^{\alpha d'} \left(\sum_{P\in\mathcal{N}}T_P\right)\star  \Gamma\nonumber \\
&= \frac{1}{|\mathcal{N}|}\left(e+\frac{2en}{d'(d'+\alpha)}\right)^{\alpha d'},\nonumber 
\end{align}
where the last equality uses the facts that $\sum_{P\in\mathcal{N}} T_P$ is a quantum comb by the definition of quantum tester and $\Gamma$ is also a quantum comb.
\end{proof}

\subsection{Technical lemmas}

\begin{lemma}\label{lemma-10260526}
For $\Gamma_P$ and $\Gamma$ defined in \cref{eq-10260541} and \cref{eq-10260542}, we have
\begin{align}
\Gamma_P\preceq \left(e+\frac{2en}{d'(d'+\alpha)}\right)^{\alpha d'}\Gamma.\label{eq-10270520}
\end{align}
\end{lemma}
\begin{proof}
Recall that $V_{P,U}=I_{d-d'}\oplus W_P U$. In this proof, for convenience, we directly write it as $V_{P,U}=(I_{d-d'}+W_{P})U$, where $I_{d-d'}:\mathbb{C}^{d-d'}\rightarrow \mathbb{C}^{d-d'}$, $W_P:\mathbb{C}^{d'}\rightarrow\mathbb{C}^{d'+\alpha}$ act on the corresponding subspaces in the decomposition in \cref{eq-10270000} and act as $0$ elsewhere, and $U$ acts non-trivially on $\mathbb{C}^{d'}$ and trivially (i.e., identity) elsewhere.

Then, note that
\begin{align}
\kett{V_{P,U}}^{\otimes n}&=\left(\kett{I_{d-d'}}+\kett{W_P U}\right)^{\otimes n}\nonumber \\
&=\sum_{i=0}^n L_i\kett{W_P U}^{\otimes i},\nonumber
\end{align}
where
\[L_i:\kett{X}^{\otimes i}\mapsto \sum_{\substack{S\subseteq [n]\\ |S|=i}}\kett{X}^{\otimes S}\otimes \kett{I_{d-d'}}^{\otimes [n]\setminus S}\]
is a linear map.
Then, note that
\begin{align}
\Gamma_P&=\E_{U\sim\mathbb{U}_{d'}}\kettbbra{V_{P,U}}{V_{P,U}}^{\otimes n}\nonumber\\
&= \E_{U\sim\mathbb{U}_{d'}}\sum_{i,j=0}^n L_i \kett{W_PU}^{\otimes i}\bbra{W_PU}^{\otimes j} L_j^\dag \nonumber\\
&= \E_{U\sim\mathbb{U}_{d'}}\sum_{i=0}^n L_i \kettbbra{W_PU}{W_PU}^{\otimes i} L_i^\dag, \nonumber
\end{align}
where we used that for $i\neq j$
\[\E_{U\sim\mathbb{U}_{d'}}\kett{W_P U}^{\otimes i}\bbra{W_P U}^{\otimes j}=\E_{u\sim\mathbb{U}_1}u^{i-j}\E_{U\sim\mathbb{U}_{d'}}\kett{W_P U}^{\otimes i}\bbra{W_P U}^{\otimes j}=0.\]
Then, we can apply \cref{lemma-10270519} to $\E_{U\sim\mathbb{U}_{d'}}\kettbbra{W_PU}{W_PU}^{\otimes i}$ for each $i\in\{0,1,\ldots,n\}$ to obtain \cref{eq-10270520}.
\end{proof}

\begin{lemma}\label{lemma-10270519}
For $\ell \geq 0$, let 
\[X_{P,\ell}=\E_{U\sim \mathbb{U}_{d'}}\kettbbra{W_P U}{W_PU}^{\otimes \ell},\qquad X_{\ell}= \E_{P\sim\Proj_{d'}(d'+\alpha)} X_{P,\ell}.\]
Then,
\[X_{P,\ell}\preceq \left(e+\frac{2e\ell}{d'(d'+\alpha)}\right)^{\alpha d'}X_\ell.\]
\end{lemma}
\begin{proof}
The case $\ell=0$ is immediate. Consider the Schur-Weyl decomposition
\[
 (\mathbb C^k)^{\otimes\ell}\cong
 \bigoplus_{\lambda\vdash_k\ell}
 \mathcal{P}_\lambda\otimes\mathcal{Q}_\lambda^k,
\]
where $\lambda\vdash_k\ell$ denotes a partition with at most $k$ rows.
Since $W_P^{\otimes\ell}$ commutes with the permutation action of $\mathfrak{S}_\ell$, we know that $W_P^{\otimes \ell}=\bigoplus_{\lambda\vdash_{d'}\ell} I_{\mathcal P_\lambda}\otimes W_{P,\lambda}$, where
$W_{P,\lambda}:\mathcal Q_\lambda^{d'}\to\mathcal Q_\lambda^{d'+\alpha}$
is an isometry. 
Let $\Pi_{\lambda,P}=W_{P,\lambda}W_{P,\lambda}^\dag$, which is thus a projector in $\mathcal{Q}_\lambda^{d'+\alpha}$ of rank $\dim \mathcal{Q}_\lambda^{d'}$.
Then, we know that
\begin{align*}
 X_{P,\ell}&=\bigoplus_{\lambda,\mu\vdash_{d'}\ell}\kettbbra{I_{\mathcal{P}_\lambda}}{I_{\mathcal{P}_\mu}}\otimes \E_{U\sim\mathbb{U}_{d'}}  \left(I_{\mathcal{Q}_\lambda^{d'+\alpha}}\otimes \texttt{q}_\lambda(U)\right) \kettbbra{W_{P,\lambda}}{W_{P,\mu}} \left(I_{\mathcal{Q}_\mu^{d'+\alpha}}\otimes \texttt{q}_\mu(U^\dag)\right)\\
 &=\bigoplus_{\lambda\vdash_{d'}\ell}
 \kettbbra{I_{\mathcal{P}_\lambda}}{I_{\mathcal{P}_\lambda}}\otimes \Pi_{\lambda,P}
 \otimes\frac{I_{\mathcal Q_\lambda^{d'}}}{\dim\mathcal Q_\lambda^{d'}},
\end{align*}
where we use Schur's lemma in the last equality. Then,
\begin{align*}
 X_\ell& =\bigoplus_{\lambda\vdash_{d'}\ell}
 \kettbbra{I_{\mathcal{P}_\lambda}}{I_{\mathcal{P}_\lambda}}\otimes I_{\mathcal Q_\lambda^{d'+\alpha}}
 \otimes\frac{I_{\mathcal Q_\lambda^{d'}}}{\dim\mathcal Q_\lambda^{d'+\alpha}},
\end{align*}
where we use that 
\[\E_{P\sim\Proj_{d'}(d'+\alpha)}\Pi_{\lambda,P}=\E_{U\sim\mathbb{U}_{d'+\alpha}}\texttt{q}_\lambda(U)\Pi_{\lambda,I_{d'}}\texttt{q}_\lambda(U)^\dag=\frac{\tr(\Pi_{\lambda,I_{d'}})}{\dim \mathcal{Q}_\lambda^{d'+\alpha}}I_{\mathcal{Q}_\lambda^{d'+\alpha}}=\frac{\dim\mathcal{Q}_\lambda^{d'}}{\dim \mathcal{Q}_\lambda^{d'+\alpha}}I_{\mathcal{Q}_\lambda^{d'+\alpha}}.\]
Consequently
\begin{equation*}
 X_{P,\ell}\preceq \max_{\lambda\vdash_{d'}\ell}
 \frac{\dim\mathcal Q_\lambda^{d'+\alpha}}{\dim\mathcal Q_\lambda^{d'}} \cdot X_\ell.
\end{equation*}
By the dimension formula of $\mathcal{Q}_\lambda^k$ (see, e.g., \cite{etingof2011introduction}), we have for $\lambda\vdash_{d'} \ell$,
\[
 \frac{\dim\mathcal Q_\lambda^{d'+\alpha}}{\dim\mathcal Q_\lambda^{d'}}
 =\prod_{i=1}^{d'}\prod_{j=1}^\alpha
 \frac{\lambda_i+d'+j-i}{d'+j-i}.
\]
%There are $\alpha d'$ factors. Their numerator has arithmetic mean$\ell/d'+(d'+\alpha)/2$. 
For the denominator, depending on grouping first by $i$ or by $j$, we can see that
\[
 \left(\prod_{i=1}^{d'}\prod_{j=1}^\alpha(d'+j-i)\right)^{1/(\alpha d')}
 \ge\max\{(d'!)^{1/d'},(\alpha!)^{1/\alpha}\}
 \ge\frac{(d'+\alpha)}{2e},
\]
where the last inequality uses $n!\geq (n/e)^n$.
Moreover, applying the arithmetic-geometric mean inequality to the numerator, we obtain 
\begin{align*}
    \left(\prod_{i=1}^{d'}\prod_{j=1}^\alpha
 (\lambda_i+d'+j-i) \right)^{1/(\alpha d')} &\le  \frac{1}{d'\alpha} \sum_{i=1}^{d'}\sum_{j=1}^\alpha
 (\lambda_i+d'+j-i) 
 = \frac{\ell}{d'}+\frac{d'+\alpha}{2}.
\end{align*}
Finally, 
\[\max_{\lambda\vdash_{d'}\ell}
 \frac{\dim\mathcal Q_\lambda^{d'+\alpha}}{\dim\mathcal Q_\lambda^{d'}}\le(e+2e\ell/(d'(d'+\alpha)))^{\alpha d'}. \]
\end{proof}

\section{Packing nets of quantum channels}\label{sec:packing-nets}
In this section, we show the construction of packing nets of quantum channels whose Stinespring dilation isometries belong to the hard family of  \cref{sec:hard-iso}. We first describe the general form of our construction in \cref{sec:general-form}, then we prove the existence of packing nets for both the   Choi trace norm and the diamond norm in \cref{sec:exist-packing-nets}.

This construction applies throughout the near-boundary regime $d_1 < rd_2 < 2d_1$. Let 
\[\alpha \coloneqq rd_2-d_1, \qquad\textup{and}\qquad \chi \coloneqq \min\{\alpha d_1, \alpha^2d_2^2\}.\]
For every $0< \eta <1$, the separations and their corresponding packing cardinality are summarized in \cref{tab:summary-packing-net}.

\begin{table}
    \centering
    \begin{tabular}{|c|c|c|}
    \hline
         & Distance & Packing cardinality\\ \hline
          Choi trace norm & $\dfrac{\sqrt{\chi}}{3e\sqrt{450}d_1} \eta$ & $\dfrac{1}{2}\left(\dfrac{1}{\eta}\right)^{2\alpha \floor{d_1/2}}$ \\ \hline
        Diamond norm &  $\dfrac{\eta}{e\sqrt{8}}$ & $\dfrac{1}{2}\left(\dfrac{1}{\eta}\right)^{2\alpha \floor{d_1/2}}$ \\
        \hline
    \end{tabular}
    \caption{Construction of packing nets.} 
    \label{tab:summary-packing-net}
\end{table}

\subsection{Construction}\label{sec:general-form}
Let $d=d_1$, $D=rd_2$, and $d' = \floor{d/2}$ so $\alpha = D-d$ satisfies $0 < \alpha <d$. Write $\calH_{\rm A_1} = \mathbb{C}^{d-d'}$, $\calH_{\rm A_2} = \mathbb{C}^{d'}$, $\calH_{\rm B} = \mathbb{C}^{d_2}$ and $\calH_{\rm anc} = \mathbb{C}^{r}$. Since $d'= \floor{d/2} \ge d/3$ and $\alpha\ge 1$, 
\begin{equation}\label{eq:dim-inequ}
    d'+\alpha \le 3d', \quad d'+\alpha \ge \frac{D}{2}, \quad \frac{dD}{6} \le d'(d'+\alpha) < \frac{3dD}{8}. 
\end{equation}

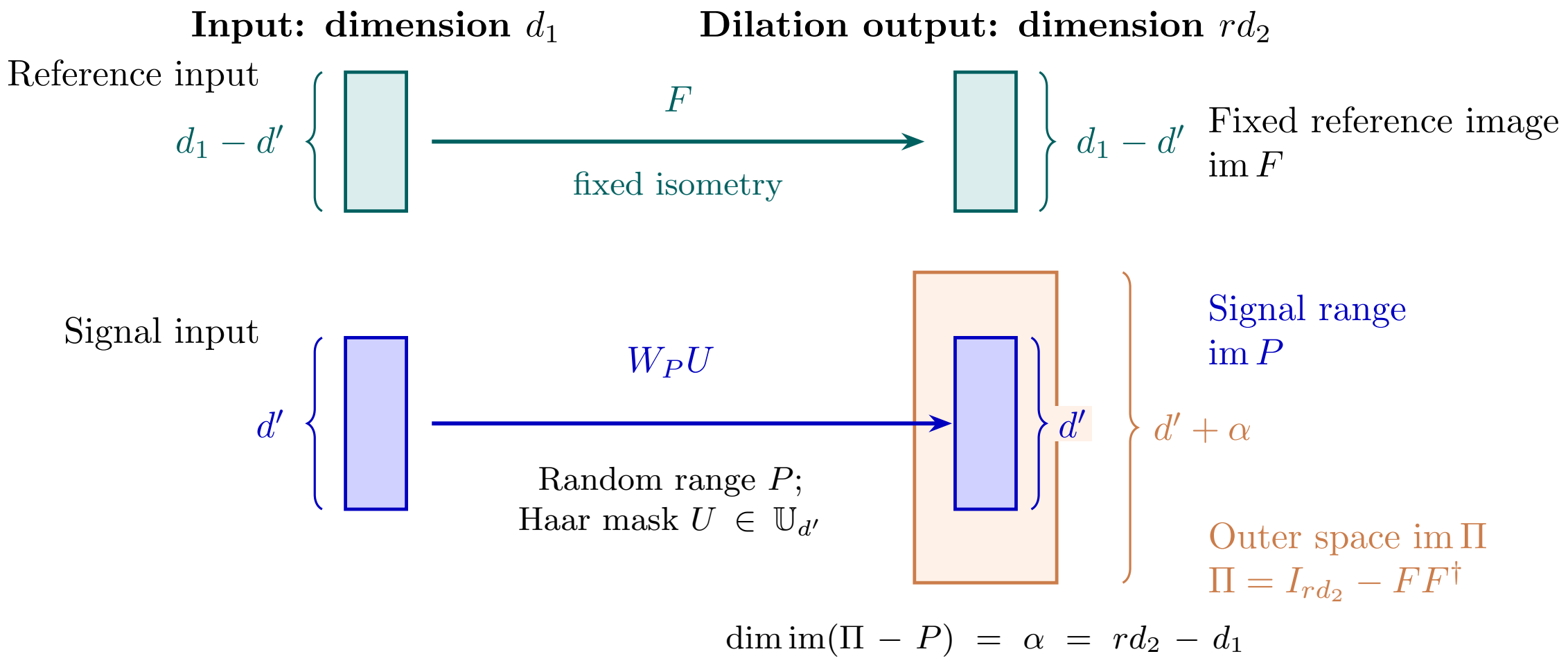
\begin{figure}[!t]
\centering
\begin{tikzpicture}[
 x=0.97cm,y=0.78cm,font=\small,
 >={Stealth[length=2.2mm]},
 reference/.style={draw=teal!75!black,fill=teal!14,line width=0.9pt},
 signal/.style={draw=blue!75!black,fill=blue!18,line width=0.9pt},
 ambient/.style={draw=Orange!80!black,fill=Orange!9,line width=0.9pt},
 dimension/.style={decorate,decoration={brace,amplitude=4pt},line width=0.6pt}
]
 %\node[font=\bfseries] at (7.4,7.25) {The construction in $d'$ notation};
 \node[font=\small\bfseries] at (3.65,6.65) {Input: dimension $d_1$};
 \node[font=\small\bfseries] at (9.65,6.65) {Dilation output: dimension $rd_2$};
 % Fixed reference map: the two narrow boxes have equal dimension.
 \node[anchor=east,align=right,text width=2.7cm] at (2.65,6.05)
   {Reference input};
 \draw[reference] (3.35,4.4) rectangle (3.95,6.1);
 \draw[dimension,teal!75!black] (3.12,4.4) -- (3.12,6.1)
   node[midway,left=6pt] {$d_1-d'$};
 \draw[->,teal!75!black,line width=1.1pt] (4.2,5.25) -- (9.05,5.25)
   node[midway,above=4pt] {$F$}
   node[midway,below=4pt,font=\footnotesize] {fixed isometry};
 \draw[reference] (9.35,4.4) rectangle (9.95,6.1);
 \draw[dimension,teal!75!black] (10.18,6.1) -- (10.18,4.4)
   node[midway,right=6pt] {$d_1-d'$};
 \node[anchor=west,align=left,text width=3.5cm] at (11.7,5.25)
   {Fixed reference image\\$\operatorname{im}F$};
 % Random signal range is drawn inside the whole complementary output space.
 \node[anchor=east,align=right,text width=2.7cm] at (2.65,2.9)
   {Signal input};
 \draw[signal] (3.35,0.75) rectangle (3.95,2.85);
 \draw[dimension,blue!75!black] (3.12,0.75) -- (3.12,2.85)
   node[midway,left=6pt] {$d'$};
 \draw[ambient] (8.95,-0.15) rectangle (10.35,3.65);
 \draw[signal] (9.35,0.75) rectangle (9.95,2.85);
 \draw[->,blue!75!black,line width=1.1pt]
   (4.2,1.8) to[out=0,in=180] (9.32,1.8);
 \node[align=center,text width=3.8cm,text=blue!75!black]
   at (6.55,2.55) {$W_PU$};
 \node[align=center,text width=3.9cm,font=\footnotesize]
   at (6.55,0.85) {Random range $P$;\\Haar mask $U\in\mathbb U_{d'}$};
 % Separate braces measure the blue inner box and the orange outer box.
 \draw[dimension,blue!75!black] (10.1,2.85) -- (10.1,0.75)
   node[midway,right=6pt,fill=Orange!9,inner sep=1pt] {$d'$};
 \draw[dimension,Orange!80!black] (11.0,3.65) -- (11.0,-0.15)
   node[midway,right=6pt,fill=white,inner sep=2pt] {$d'+\alpha$};
 \node[anchor=west,align=left,text width=3.5cm,text=blue!75!black]
   at (11.7,2.95) {Signal range\\$\operatorname{im}P$};
 \node[anchor=west,align=left,text width=3.5cm,text=Orange!80!black]
   at (11.7,0.1) {Outer space $\operatorname{im}\Pi$\\$\Pi=I_{rd_2}-FF^\dag$};
 \node[align=center,font=\footnotesize,text width=6.5cm]
   at (9.65,-0.85)
   {$\dim\operatorname{im}(\Pi-P)=\alpha=rd_2-d_1$};
\end{tikzpicture}
\caption{The  construction in the notation $d'=\lfloor d_1/2\rfloor$ and $\alpha=rd_2-d_1$.
The teal reference input and image have dimension $d_1-d'$.
The blue signal input and range $\operatorname{im}P$ have dimension $d'$.
The orange outer space
$\operatorname{im}\Pi=(\operatorname{im}F)^\perp
\subseteq\mathcal H_{\mathrm B}\otimes\mathcal H_{\mathrm{anc}}$
has dimension $d'+\alpha$; its unused subspace
$\operatorname{im}(\Pi-P)$ has dimension $\alpha$.}
\label{fig:range-construction-general}
\end{figure}

Our construction uses a reference map and a signal map. The reference map is balanced and its existence follows from \cite{chenoufkir2026incoherent}. 
\begin{lemma}[Balanced reference]\label{lem:balanced-ref}
There exists an isometry $F : \calH_{\rm A_1} \rightarrow \calH_{\rm B} \otimes \calH_{\rm anc}$  satisfying 
\begin{equation*}
    \tr_B(FF^\dagger) \succeq \lambda_0 I_r, \quad \lambda_0 \coloneqq  \frac{d-d'}{2r}. 
\end{equation*}
    
\end{lemma}
\begin{proof}
    Apply \cite[Lemma 4.2]{chenoufkir2026incoherent} with input dimension, $d-d'$, output dimension $d_2$ and Kraus rank $r$. Its assumptions hold: We have that $rd_2 < 2d_1$ and $d_2\ge 2$ so, $r<d_1 \le 2 (d-d') \le (d-d')d_2$. Moreover, $d-d' \le d_1 \le rd_2$. 
\end{proof}
Fix such an isometry $F$ and define $ \Pi = I_D - FF^\dagger$ which is a projector of rank $D - (d-d') = \alpha + d'$. Choose a fixed unitary identification $\mathbb{C}_D \cong \calH_{\rm B} \otimes  \calH_{\rm anc} $ sending the  first $d-d'$ canonical basis vectors to the columns of $F$. For each $P \in \Proj_{d'}(\mathrm{im} \Pi)$, fix an isometry $W_P:\mathbb{C}^{d'}\to \mathrm{im}\Pi$ with $W_PW_P^\dagger = P$.  For  $U\in \mathbb{U}(d')$, define 
\begin{equation}\label{eq:family}
    V_{P, U} \coloneqq \begin{pmatrix}
        F  & W_P U
    \end{pmatrix}, \quad \calE_{P,U} \coloneqq \tr_{\rm anc}\circ \calV_{P,U}
\end{equation}
See \cref{fig:range-construction-general} for an illustration of our construction. Let $\mu$ be the invariant probability measure on  $\Proj_{d'}(\mathrm{im} \Pi)$, obtained by conjugating a fixed projector by a Haar unitary. \cref{thm:hard-isom-disc} applies after a fixed change of output basis.

\begin{proposition}[Valid construction]\label{prop:hard-family}
Every $V_{P,U}$ in \cref{eq:family} is an isometry, and
$\mathcal E_{P,U}$ is a channel with Kraus rank exactly $r$.
\end{proposition}
\begin{proof}
    We have that $F^\dagger W_P=0$ so 
    \begin{align}
        V_{P,U}^\dagger V_{P,U} &= \begin{pmatrix} F^\dagger F & F^\dagger W_PU \\ U^\dagger W_P^\dagger F & U^\dagger W_P^\dagger W_P U \end{pmatrix} = \begin{pmatrix} I_{d_1-d'} & 0 \\ 0 & I_{d'} \end{pmatrix} =I_d. 
    \end{align}
By its definition in \cref{eq:family}, the channel $\calE_{P, U}$ has Kraus rank at most $r$. Moreover, we can write $A_{V_{P,U}}$ as follows
\begin{equation*}
    A_{V_{P,U}} = \begin{pmatrix}
        A_{F} \\ A_{W_PU}
    \end{pmatrix}.
\end{equation*}
By \cref{lem:balanced-ref}, we have that 
\begin{equation*}
    A_{F}^\dagger A_{F} =  \left(\tr_B(FF^\dagger)\right)^{\top} \succeq \lambda_0 I_r.
\end{equation*}
Hence 
\begin{equation*}
    A_{V_{P,U}}^\dagger A_{V_{P,U}}  = A_{F}^\dagger A_{F} + A_{W_PU}^\dagger A_{W_PU} \succeq \lambda_0 I_r\succ 0. 
\end{equation*}
Therefore $A_{V_{P,U}}$ has rank $r$, and 
\begin{equation*}
    \rank C_{\calE_{P,U}} = \rank\left( A_{V_{P,U}} A_{V_{P,U}}^\dagger \right) = r.
\end{equation*}
Finally, the channel $\calE_{P, U}$ has Kraus rank exactly $r$.
\end{proof}

\subsection{Existence}\label{sec:exist-packing-nets}
We restrict the signal ranges $P$ of our construction (\cref{eq:family}) to the good set $\calG$ defined as 
\begin{equation}
    \calG = \left\{ P \in  \Proj_{d'}(\mathrm{im} \Pi) : \tr_B(\Pi - P) \preceq \kappa  I_r   \right\}
\end{equation}
where $ \kappa = C_0(\frac{\alpha}{r} +\frac{1}{d_2})$ and $C_0 = 450$. 

\begin{lemma}[Balanced Haar signals]\label{lem:good-signals}
The set $\mathcal G$ is compact, $\mu(\mathcal G)\geq1/2$, and
\begin{equation}\label{eq:K-comparison}
 \frac \chi{16C_0}\leq\frac{\lambda_0\alpha^2}{\kappa}
                     \leq\frac \chi{2C_0}.
\end{equation}
\end{lemma}
The proof of \cref{lem:good-signals} is deferred to
\cref{sec:packing-balanced-signals}.

We now show the existence of the packing net. 
\begin{theorem}[Existence of packing net]\label{thm:packing}
For each $0<\eta<1$, there is a finite $\mathcal N\subseteq\mathcal G$
such that
\begin{equation}\label{eq:packing-size}
 |\mathcal N|\geq\tfrac12(1/\eta)^{2\alpha d'},\qquad
 \|P-Q\|_1>\frac{4\alpha d'}{e(d'+\alpha)}\eta
 \quad(P\ne Q\in\mathcal N).
\end{equation}
Let $\mathcal O_P=\{\mathcal E_{P,U}:U\in\mathbb U_{d'}\}$.
For distinct $P,Q\in\mathcal N$, the separation holds uniformly:
\begin{align}
 \inf_{\mathcal E\in\mathcal O_P,\mathcal F\in\mathcal O_Q}
   \frac1{d_1}\|C_{\mathcal E}-C_{\mathcal F}\|_1
 &>\frac{\sqrt \chi}{3e\sqrt{C_0}\,d_1}\eta,
                       \label{eq:packing-choi}\\
 \inf_{\mathcal E\in\mathcal O_P,\mathcal F\in\mathcal O_Q}
   \|\mathcal E-\mathcal F\|_\diamond
 &>\frac\eta{e\sqrt8}.\label{eq:packing-diamond}
\end{align}
\end{theorem}
\begin{proof}
Let  $s=\frac{4\alpha d'\eta}{e(d'+\alpha)}>0$. Choose $\calN$ to be  a maximal subset of
$\mathcal G$ with pairwise distances strictly greater than $s$ in trace
norm. Such a maximal packing exists and is finite because $\calG$ is compact~\cite{Ver18}.
By maximality of the packing $\calN$, the closed radius-$s$ balls centered at elements of $\calN$ cover $\mathcal G$, so
\cref{lem:good-signals,lem:haar-small-ball} give
$$
 \frac12\leq\mu(\mathcal G)
 \leq\sum_{P\in\mathcal N}\mu\{Q:\|P-Q\|_1\leq s\}
 \leq|\mathcal N|\eta^{2\alpha d'}.
$$
This proves \cref{eq:packing-size}.

Let $P, Q \in \calN$. 
By \cref{lem:unif-sep-choi}, we have that 
$$
 \inf_{\mathcal E\in\mathcal O_P,\mathcal F\in\mathcal O_Q}
   \frac1{d_1}\|C_{\mathcal E}-C_{\mathcal F}\|_1 \ge \frac{4\alpha d'}{ed_1(d'+\alpha)}
       \sqrt{\frac{\lambda_0}{\kappa}}\eta
 \geq\frac{\sqrt \chi}{3e\sqrt{C_0}\,d_1}\eta,
$$
where we used $d'/(d'+\alpha)\geq1/3$ (\cref{eq:dim-inequ}) and
$\alpha\sqrt{\lambda_0/\kappa}\geq\sqrt \chi/(4\sqrt{C_0})$ (\cref{lem:good-signals}).
Since $\rank(P-Q)\leq2\min\{\alpha,d'\}$,
$$
 \|P-Q\|_{\mathrm{op}}
 \geq\frac{\|P-Q\|_1}{2\min\{\alpha,d'\}}
 >\frac{2\max\{\alpha,d'\}}{e(\alpha+d')}\eta
 \geq\frac\eta e.
$$
Therefore, by \cref{lem:unif-sep-dia} we obtain 
\begin{equation*}
    \inf_{\mathcal E\in\mathcal O_P,\mathcal F\in\mathcal O_Q}
   \|\mathcal E-\mathcal F\|_\diamond
 \ge \frac{1}{\sqrt{8}}  \|P-Q\|_{\mathrm{op}} > \frac{\eta}{e\sqrt{8}}. 
\end{equation*}
\end{proof}

\subsection{Balanced Haar signals: proof of \cref{lem:good-signals}}\label{sec:packing-balanced-signals}

We first establish the Gaussian estimates needed to prove \cref{lem:good-signals}.
\begin{lemma}[Gaussian operator-norm and least-singular-value estimates]\label{lem:gaussian-estimates}
If $G\in\mathbb C^{a\times b}$ is centered circular complex Gaussian
with vectorized covariance at most identity, then
\begin{equation}\label{eq:gaussian-op}
 \Pr\{\|G\|_{\mathrm{op}}>3\sqrt{a+b}\}\leq e^{-(a+b)}.
\end{equation}
If $G_0\in\mathbb C^{m\times\alpha}$ has independent $\calC\calN(0,1)$
entries and $m\geq125\alpha$, then
\begin{equation}\label{eq:gaussian-min}
 \Pr\{\sigma_{\min}(G_0)\leq\sqrt m/5\}\leq2e^{-m/16}.
\end{equation}
Here a standard complex Gaussian has independent real and imaginary
parts of variance $1/2$.
\end{lemma}
\begin{proof}

Write $C=\operatorname{Cov}(\operatorname{vec}G)\preceq I_{ab}$.
Choose an independent centered circular Gaussian matrix $G'$ with
vectorized covariance $I_{ab}-C$.
Then $H=G+G'$ is circular Gaussian with vectorized covariance $I_{ab}$,
so its entries are independent $\calC\calN(0,1)$ variables.
Since $\E[H\mid G]=G$, conditional Jensen gives
$\E\|G\|_{\mathrm{op}}\leq\E\|H\|_{\mathrm{op}}$.
Write $H=(X+\mathrm iY)/\sqrt2$, where $X,Y$ have independent
$\calN(0,1)$ entries. Applying \cite[Theorem~7.3.1]{Ver18} to $X$ and $Y$ gives
$$
 \E\|G\|_{\mathrm{op}}
 \leq\frac{\E\|X\|_{\mathrm{op}}+\E\|Y\|_{\mathrm{op}}}{\sqrt2}
 \leq\sqrt2(\sqrt a+\sqrt b)\leq2\sqrt{a+b}.
$$
For concentration, represent
$\operatorname{vec}G=C^{1/2}(x+\mathrm i y)/\sqrt2$ with independent
standard real Gaussian vectors $x,y\in\mathbb R^{ab}$.
The function $(x,y)\mapsto\|G\|_{\mathrm{op}}$ is
$1/\sqrt2$-Lipschitz because $\|C^{1/2}\|_{\mathrm{op}}\leq1$
and operator norm is bounded by Frobenius norm.
Thus Gaussian concentration
\cite[Theorem~5.6]{boucheron2013concentration} gives
$$
 \Pr\{\|G\|_{\mathrm{op}}>\E\|G\|_{\mathrm{op}}+t\}\leq e^{-t^2}.
$$
Taking $t=\sqrt{a+b}$ proves \cref{eq:gaussian-op}.
The least-singular-value estimate \cref{eq:gaussian-min} is precisely
\cite[Lemma~3.9]{chenoufkir2026incoherent}, with its row and column
dimensions set to $m$ and $\alpha$, respectively.
\end{proof}

We now apply these estimates to the good set $\mathcal G$.

\begin{lemma}[Restatement of \cref{lem:good-signals}]
The set $\mathcal G$ is compact, $\mu(\mathcal G)\geq1/2$, and
\begin{equation}\label{eq:K-comparison-r}
 \frac \chi{16C_0}\leq\frac{\lambda_0\alpha^2}{\kappa}
                     \leq\frac \chi{2C_0}.
\end{equation}
\end{lemma}
\begin{proof}
    Since the map $P\mapsto \tr_B(\Pi- P)$ is continuous, $\calG$ is compact.

    Let $Q = \Pi - P$ be a Haar random projector of rank $\alpha$ in $\mathrm{im} \Pi$. If $\alpha \ge (d'+\alpha)/125$, then  $\tr_B Q \preceq d_2 I_r\preceq \frac{250\alpha}{r} I_r $ because $\alpha \ge (d'+\alpha)/125 \ge rd_2/250$ (\cref{eq:dim-inequ}). So, in this case, every $P$ belongs to $\calG$. 

    Suppose that $\alpha < (d'+\alpha)/125$. Let $Z\in \mathbb{C}^{rd_2\times \alpha}$ have independent $\calC\calN(0,1)$ entries. Let 
    \begin{equation*}
        G= \Pi Z, \quad  N= G(G^\dagger G)^{-1/2}, \quad Q=NN^\dagger. 
    \end{equation*}
By unitary invariance of the Gaussian distribution, $G$ has independent $\calC\calN(0,1)$ coordinates in an orthonormal basis of $\mathrm{im}\Pi$ and $Q$ is a Haar-distributed projector of rank $\alpha$ on $\rm{im}\Pi$; hence $Q$ has the same law as $\Pi-P$ for $P\sim\mu$, since $\rank\Pi=d'+\alpha$.

% Recall that we can view  $G : \mathbb{C}^\alpha \rightarrow \calH_{\rm B} \otimes \calH_{\rm anc}$ so it can be written as $G = \sum_{e=1}^r G_e\otimes \ket{e}_{\rm anc}$ and we can define 
% \begin{align}
%      A_G= \sqrt{\frac{r}{\alpha}} \begin{pmatrix}
%         \kett{G_1} & \kett{G_2} & \cdots & \kett{G_r}  
%     \end{pmatrix} \in \mathbb{C}^{d_2\alpha \times r}.
% \end{align}
% We have that 
Using the reshaping $A_G\in\mathbb C^{d_2\alpha\times r}$ from \cref{eq:Kraus} and its Gram-matrix identity in \cref{eq:reshaping-identities}, we have
\begin{align*}
    \tr_B\left(Q \right) &= \tr_B\left(G(G^\dagger G)^{-1}G^\dagger\right)
    \\&\preceq \frac{1}{\sigma_{\min}(G)^2} \tr_B\left(GG^\dagger\right)
    \\&= \frac{1}{\sigma_{\min}(G)^2}\left(A_{G}^\dagger A_{G}\right)^{\top}.
\end{align*}
    The entries of $ A_{G}$ are a coordinate
permutation of those of $G$. Its vectorized covariance is therefore
a coordinate permutation of $I_\alpha\otimes\Pi\preceq I$. Hence, by \cref{lem:gaussian-estimates}, with probability at least $1/2$ ($\alpha+d'\ge 126$, $r+d_2\alpha\ge 3$ so by union bound, the error probability is at most $e^{-3} + 2e^{-126/16}<1/2$), we have that 
\begin{equation*}
    \sigma_{\min}(G)  \ge  \frac15\sqrt{\alpha+d'}, \quad \|A_{G}\|_{\rm op} \le 3 \sqrt{r+d_2\alpha}. 
\end{equation*}
These inequalities imply
\begin{align*}
    \tr_B\left(Q \right) &\preceq \frac{ \|A_{G}^\dagger\|_{\rm op}^2}{\sigma_{\min}(G)^2}I_r 
    \\& \preceq 225 \frac{r+d_2\alpha}{\alpha +d'} I_r
    \\& \preceq 450 \left( \frac{1}{d_2} +\frac{\alpha}{r} \right) I_r \preceq \kappa I_r,
\end{align*}
using $\alpha + d' \ge rd_2/2$ (\cref{eq:dim-inequ}).

Finally,
$$
 \frac{\lambda_0\alpha^2}{\kappa}
 =\frac{(d-d')\alpha^2}{2C_0(\alpha+r/d_2)}.
$$
For $x,y>0$, since $\max\{x,y\}\le x+y\le 2\max\{x,y\}$, we deduce that
$\tfrac12\min\{x,x^2/y\}\leq x^2/(x+y)\leq\min\{x,x^2/y\}$.
Here $d/2\leq d-d'\leq d$, and
$d/d_2^2<r/d_2<2d/d_2^2$. Substituting these inequalities gives
the lower bound $\chi/(16C_0)$ and upper bound $\chi/(2C_0)$ in
\cref{eq:K-comparison-r}.
\end{proof}

\subsection{Separation uniformly over the Haar mask} 

In this section we show separation guarantees for both   Choi trace norm and diamond norm given the separation of the signal ranges and uniformly over the Haar masks. 

We use the same perturbation notation as in the upper bound. For
$V=V_{P,U}$ and $\widehat V=V_{Q,U'}$, put
$$
 z=\tr(V^\dag\widehat V)/d_1,\qquad
 \Deltain=V^\dag\widehat V-zI_{d_1},\qquad
 \Deltaout=(I-VV^\dag)\widehat V.
$$
Then $\widehat V=zV+V\Deltain+\Deltaout$.
\begin{lemma}[Range displacement]\label{lem:range-displacement}
For $P,Q,U,U'$ and $\Deltaout$ as above, let
$N_P:\mathbb C^\alpha\to\operatorname{im}\Pi$ be an isometry with
$N_PN_P^\dag=\Pi-P$. Then
\begin{equation}\label{eq:range-displacement}
 \begin{gathered}
 \Deltaout=[\,0\ \ N_PN_P^\dag W_QU'\,],\qquad
 \|\Deltaout\|_1=\tfrac12\|P-Q\|_1,\\
 \|\Deltaout\|_{\mathrm{op}}=\|P-Q\|_{\mathrm{op}},\qquad
 \|\Deltaout\|_{\mathrm F}^2=\tfrac12\|P-Q\|_{\mathrm F}^2.
 \end{gathered}
\end{equation}
Moreover, $\rank(P-Q)\leq2\min\{\alpha,d'\}$.
\end{lemma}
\begin{proof}
    The first equality is deduced from $VV^\dagger = FF^\dagger +P$. For the remaining ones, we observe that the nonzero singular values of $N_P^\dagger W_Q$ are the sines of the principal angles between $P$ and $Q$, and the nonzero singular values of $P-Q$ consist of each such sine twice. 
\end{proof}

\begin{lemma}\label{lem:unif-sep-choi}
    Let $P \in \calG$, $Q\in \Proj_{d'}(\mathrm{im} \Pi)$ and $U, U' \in \mathbb{U}(d')$. We have that
    \begin{align}
         \frac1{d_1}\|C_{\mathcal E_{P,U}}-C_{\mathcal E_{Q,U'}}\|_1 &\geq\frac1{d_1}\sqrt{\frac{\lambda_0}{\kappa}}\|P-Q\|_1.
    \end{align}
\end{lemma}
\begin{proof}
    Let $W = W_PU$ and $W' = W_QU'$. We have that $
    A_{V_{P,U}} = \begin{pmatrix}
        A_{F} \\ A_{W}
    \end{pmatrix}$   and $A_{V_{Q, U'}} = \begin{pmatrix}
        A_{F} \\ A_{W'}
    \end{pmatrix}$ so by \cref{eq:Kraus-Choi}, the $(0,1)$ block of the Choi operator  is $A_FA_W^\dagger$ for the channel $\calE_{P,U}$   and $A_FA_{W'}^\dag$ for the channel $\calE_{Q,U'}$.   

    By \cref{lem:balanced-ref,eq:reshaping-identities}, we have that 
    \begin{equation}
        A_F^\dagger A_F = \left(\tr_B FF^\dagger\right)^\top \succeq \lambda_0 I_r.
    \end{equation}
    For a contraction $Y\in\mathbb C^{\alpha\times d'}$, and since  $P\in\mathcal G$, we have that using \cref{eq:reshaping-identities}
$$
 A_{N_PY}^\dag A_{N_PY}
 =\bigl(\tr_{\mathrm B}N_PYY^\dag N_P^\dag\bigr)^{\top}
 \preceq\bigl(\tr_{\mathrm B}(\Pi-P)\bigr)^{\top}
 \preceq\kappa I_r.
$$
Define
$$
 X=\sqrt{\frac{\lambda_0}{\kappa}}\,
       A_F(A_F^\dag A_F)^{-1}A_{N_PY}^\dag,
 \qquad H=\begin{pmatrix}0&X\\X^\dag&0\end{pmatrix}.
$$
The preceding bounds imply
$$
 \|X\|_{\mathrm{op}}
 \leq\sqrt{\frac{\lambda_0}{\kappa}}
      \frac{1}{\sqrt{\lambda_0}}\sqrt{\kappa}=1,
 \qquad \|H\|_{\mathrm{op}}\leq1.
$$
Using the inner-product identity in \cref{eq:reshaping-inner-product} and $N_P^\dag W=0$, we obtain
\begin{align*}
 \frac1{d_1}\tr\bigl(H(C_{\mathcal E_{P,U}}-C_{\mathcal E_{Q,U'}})\bigr)
 &=\frac{2}{d_1}\sqrt{\frac{\lambda_0}{\kappa}}\operatorname{Re} \tr\bigl(A_{N_PY}^\dagger(A_W-A_{W'})      \bigr)
 \\
 &=\frac2{d_1}\sqrt{\frac{\lambda_0}{\kappa}}
       \operatorname{Re}\tr\bigl(Y^\dag N_P^\dag(W-W')\bigr)\\
 &=-\frac2{d_1}\sqrt{\frac{\lambda_0}{\kappa}}
       \operatorname{Re}\tr\bigl(Y^\dag N_P^\dag W'\bigr).
\end{align*}
 Therefore by \cref{lem:range-displacement} we deduce that 
 \begin{align*}
      \frac1{d_1}\|C_{\mathcal E_{P,U}}-C_{\mathcal E_{Q,U'}}\|_1
 &\geq \max_{Y : \|Y\|_{\rm op}\le 1} -\frac2{d_1}\sqrt{\frac{\lambda_0}{\kappa}}
       \operatorname{Re}\tr\bigl(Y^\dag N_P^\dag W'\bigr)
 \\&=\frac2{d_1}\sqrt{\frac{\lambda_0}{\kappa}}
       \|N_P^\dag W'\|_1
 =\frac1{d_1}\sqrt{\frac{\lambda_0}{\kappa}}\|P-Q\|_1.
 \end{align*}

\end{proof}
\begin{lemma}\label{lem:unif-sep-dia}
    Let $P, Q \in \Proj_{d'}(\mathrm{im} \Pi)$ and $U, U' \in \mathbb{U}(d')$. We have that
    \begin{align}
      \|\mathcal E_{P,U}-\mathcal E_{Q,U'}\|_\diamond
 &\geq\frac1{\sqrt8}\|P-Q\|_{\mathrm{op}}.
    \end{align}
\end{lemma}
\begin{proof}
Let $W=W_PU$ and $W'=W_QU'$, and let 
$x\in\mathbb C^{d'}$ be a unit input such that
$t=(W-W')x$ satisfies $\|t\|_2=\|W-W'\|_{\mathrm{op}}$.
If $t=0$, then $W=W'$ and hence $P=Q$, so the claim is immediate.
Otherwise, let
$S_{a,b}\ket j=e^{2\pi\mathrm i bj/d_2}\ket{j+a\bmod d_2}$,
$a,b\in\{0,\ldots,d_2-1\}$, be the Weyl unitaries.
Their twirling identity and \cref{lem:balanced-ref} imply
$$
 \frac1{d_2^2}\sum_{a,b}
 \|F^\dag(S_{a,b}\otimes I_r)t\|_2^2
 =\left\langle t\middle| \left(\frac{I_{\mathrm B}}{d_2}
        \otimes\tr_{\mathrm B}(FF^\dag)\right)t\right\rangle
 \geq\frac{\lambda_0}{d_2}\|t\|_2^2.
$$
Choose $S=S_{a,b}$ attaining at least this average, put
$h=F^\dag(S\otimes I_r)t$, and take the unit  input
$y=h/\|h\|_2\in\mathbb C^{d_1-d'}$.

For an auxiliary qubit $R$, we use the normalized input
$$
 \ket\psi=\frac{\ket{y}_{\mathrm A,0}\ket0_R+
                       \ket{x}_{\mathrm A,1}\ket1_R}{\sqrt2}.
$$
Let
$$
 M=\bigl((\mathcal E_{P,U}-\mathcal E_{Q,U'})\otimes\mathrm{id}_R\bigr)
           (\ketbra\psi\psi).
$$
We have that 
\begin{equation*}
   \bigl( \mathcal E_{P,U}\otimes\mathrm{id}_R\bigr)
           (\ketbra\psi\psi) = \frac12\begin{pmatrix}\tr_{\mathrm{anc}} F\proj{y}{y} F^\dag & \tr_{\mathrm{anc}} F\ket{y}\bra{x} W^\dag
           \\\tr_{\mathrm{anc}}W\ket{x}\bra{y}F^\dag  &\tr_{\mathrm{anc}}W \proj{x}{x}W^\dag
           \end{pmatrix}
\end{equation*}
Hence 
\begin{equation*}
   M=\bigl((\mathcal E_{P,U}-\mathcal E_{Q,U'})\otimes\mathrm{id}_R\bigr)
           (\ketbra\psi\psi) = \frac12\begin{pmatrix}0 & T^\dag
           \\T  &*
           \end{pmatrix}
\end{equation*}
with $T:=\tr_{\mathrm{anc}}\ketbra{t}{Fy}.$
Since $S$ is a unitary, we have
\begin{equation}\label{eq:diam-2}
     \|T\|_1\geq|\tr(ST)|
 =|\langle Fy|(S\otimes I_r)t\rangle|
 =\|h\|_2\geq\sqrt{\frac{\lambda_0}{d_2}}\|t\|_2.
\end{equation}
Writing $Z_R=\ketbra0{0}-\ketbra1{1}$ and
$\mathcal Z=I_{\mathrm B}\otimes Z_R$, we obtain
\begin{equation}\label{eq:diam-1}
     \|T\|_1
 =\frac12\|M-\mathcal ZM\mathcal Z\|_1
 \leq\|M\|_1
 \leq\|\mathcal E_{P,U}-\mathcal E_{Q,U'}\|_\diamond.
\end{equation}
Finally, $(I-P)W=0$ and
\cref{lem:range-displacement} give
\begin{equation}\label{eq:diam-3}
     \|t\|_2=\|W-W'\|_{\mathrm{op}}
 \geq\|(I-P)W'\|_{\mathrm{op}}
 =\|P-Q\|_{\mathrm{op}},\qquad
 \frac{\lambda_0}{d_2}=\frac{d_1-d'}{2rd_2}>\frac18,
\end{equation}
where we use  $d'\leq d_1/2$ and $rd_2<2d_1$.
Combining \cref{eq:diam-1,eq:diam-2,eq:diam-3} yields
$$
 \|\mathcal E_{P,U}-\mathcal E_{Q,U'}\|_\diamond
 \geq\|T\|_1
 \geq\sqrt{\frac{\lambda_0}{d_2}}\|t\|_2
 \geq\frac1{\sqrt8}\|P-Q\|_{\mathrm{op}},
$$
as required.
\end{proof}

\subsection{Haar small ball estimates}
In this section, we give  ball estimates  under the measure $\mu$ in $\Proj_{d'}(\mathrm{im}\Pi)$. Before that, we list several facts needed for the proof. 

\begin{fact}[Projector distance in terms of the lower block]\label{item:small-ball-distance}
Let $d',\alpha\geq1$ be integers, let
$W=\binom AB\in\mathbb C^{(d'+\alpha)\times d'}$ satisfy
$W^\dag W=I_{d'}$, with $A\in\mathbb C^{d'\times d'}$ and
$B\in\mathbb C^{\alpha\times d'}$, and put
$P_0=\operatorname{diag}(I_{d'},0_\alpha)$ and $Q=WW^\dag$.
The two-subspace decomposition \cite{halmos1969subspaces} and the
principal-angle singular-value characterization
\cite{bjorck1973angles} give the following identity
$$
 \|P_0-Q\|_1=2\|B\|_1.
$$
%Indeed, the nonzero eigenvalues of $P_0-Q$ occur in pairs$\pm\sin\theta_j$, while the nonzero singular values of $B$ are$\sin\theta_j$, where the $\theta_j$ are the principal angles.This is also the identity in \cref{eq:range-displacement}.

\end{fact}

\begin{fact}[Operator-ball and Euclidean-ball volumes]\label{item:small-ball-volumes}
Let $d',\alpha\geq1$ be integers and put $n=\alpha d'$.
In $\mathbb C^{\alpha\times d'}$, equipped with real Lebesgue volume
for the inner product $\operatorname{Re}\tr(X^\dag Y)$, define
$\mathcal B_{\mathrm{op}}=\{B:\|B\|_{\mathrm{op}}<1\}$ and
$\mathcal B_{\mathrm F}=\{B:\|B\|_{\mathrm F}\leq1\}$.
Hua's matrix-ball integral \cite{hua1963harmonic}, as stated in
\cite[Section~2.1, equation~(2.1), at $\lambda=0$]{neretin2015matrix},
and the Euclidean unit-ball formula in real dimension $2n$ give
$$
 V_{\mathrm{op}}:=\operatorname{Vol}(\mathcal B_{\mathrm{op}})
 =\pi^n\prod_{i=1}^{\alpha}\frac{(i-1)!}{(d'+i-1)!},
 \qquad
 V_{\mathrm F}:=\operatorname{Vol}(\mathcal B_{\mathrm F})
 =\frac{\pi^n}{n!}.
$$

\end{fact}
\begin{fact}[Polarity and volume product]\label{item:small-ball-polarity}
Let $d',\alpha\geq1$ be integers and put $n=\alpha d'$.
Equip $\mathbb C^{\alpha\times d'}$ with the real inner product
$\langle X,Y\rangle=\operatorname{Re}\tr(X^\dag Y)$ and its Lebesgue
volume. Define
$\mathcal B_{\mathrm{op}}=\{B:\|B\|_{\mathrm{op}}<1\}$,
$\mathcal B_*=\{B:\|B\|_1\leq1\}$, and
$\mathcal B_{\mathrm F}=\{B:\|B\|_{\mathrm F}\leq1\}$, and write
$V_{\mathrm{op}}=\operatorname{Vol}(\mathcal B_{\mathrm{op}})$ and
$V_{\mathrm F}=\operatorname{Vol}(\mathcal B_{\mathrm F})$.
For a convex body $K$, its polar is
$K^\circ=\{Y:\langle X,Y\rangle\leq1\text{ for every }X\in K\}$.
Trace-norm duality
\cite[Section~1.1, equation~(1.173)]{watrous2018theory}
makes $\mathcal B_*$ the polar of the closed
operator-norm unit ball. The symmetric Blaschke--Santal\'o inequality
\cite[Theorem~1, Section~3]{bianchi2015fourier} therefore gives
$$
 \operatorname{Vol}(\mathcal B_*)V_{\mathrm{op}}\leq V_{\mathrm F}^2.
$$
The original volume-product inequality is due to
Blaschke \cite{blaschke1917affine} in dimensions $2$ and $3$ and
Santal\'o \cite{santalo1949invariante} in arbitrary dimension.
For any $t\geq0$, real dimension $2n$ gives
$\operatorname{Vol}(t\mathcal B_*)=t^{2n}\operatorname{Vol}(\mathcal B_*)$.
The open and closed operator balls have the same volume: their
boundary lies in the zero set of the nonzero real polynomial
$\det(I_{d'}-B^\dag B)$.
\end{fact}
\begin{fact}[Haar  block law]\label{lem:haar-block-law}
Let $d',\alpha\geq1$ be integers, and let $W$ be a Haar isometry from $\mathbb C^{d'}$ to
$\mathbb C^{d'+\alpha}$, and write
$W=\left(\begin{smallmatrix}A\\B\end{smallmatrix}\right)$ with
$A\in\mathbb C^{d'\times d'}$, $B\in\mathbb C^{\alpha\times d'}$.
Then $B$ is uniformly distributed with respect to Lebesgue measure on
$\mathcal B_{\mathrm{op}}=\{B:\|B\|_{\mathrm{op}}<1\}$.
\end{fact}
\begin{proof}
Realize $W$ as the first $d'$ columns of a Haar unitary of size
$N=d'+\alpha$. By invariance under row permutations, $B$ has the law of
an $\alpha\times d'$ corner of that unitary.
For an $n\times m$ corner $Q$, with $m\leq n$ and $N\geq m+n$,
the density with respect to Lebesgue measure is proportional to
$\det(I_m-Q^\dag Q)^{N-m-n}$ on the operator-norm unit ball
\cite[Section~3, equation~(38)]{fyodorov2007colour}.
The rectangular-block formula is attributed there to
\cite[Section~3, equations~(3.10)--(3.12), at $\beta=2$]{forrester2006conductance}.
If $\alpha\geq d'$, apply this formula with $n=\alpha$ and $m=d'$.
If $\alpha<d'$, apply it to $B^\dag$, which is a $d'\times\alpha$
corner of a Haar unitary after a row permutation of the adjoint unitary.
In either case $N=m+n$, so the density is constant.
Taking adjoints preserves Lebesgue measure and the operator norm,
which proves the claim in both cases.
\end{proof}

\begin{lemma}[Haar small balls]\label{lem:haar-small-ball}
    For every $P\in\Proj_{d'}(\mathrm{im}\Pi)$ and $s\geq0$,
$$
 \mu\{Q:\|P-Q\|_1\leq s\}
 \leq \min\left\{1, 
       \left(\frac{e(d'+\alpha)s}{4\alpha d'}\right)^{2\alpha d'}\right\}.
$$
\end{lemma}
\begin{proof}
By unitary invariance of $\mu$ and of the trace norm, it suffices to
take $P=P_0:=\operatorname{diag}(I_{d'},0_\alpha)$.
Let $W=\binom AB:\mathbb C^{d'}\longrightarrow\mathbb C^{d'+\alpha}$
be a Haar-distributed isometry, with $A\in\mathbb C^{d'\times d'}$
and $B\in\mathbb C^{\alpha\times d'}$. Then $Q=WW^\dag$ has law $\mu$.
Put $n=\alpha d'$ and $t=s/2$.
In $\mathbb C^{\alpha\times d'}$, let
$\mathcal B_{\mathrm{op}}=\{B:\|B\|_{\mathrm{op}}<1\}$,
$\mathcal B_*=\{B:\|B\|_1\leq1\}$, and
$\mathcal B_{\mathrm F}=\{B:\|B\|_{\mathrm F}\leq1\}$.
Write $V_{\mathrm{op}}$ and $V_{\mathrm F}$ for their operator-ball
and Frobenius-ball volumes, respectively, using real Lebesgue volume
for the inner product $\operatorname{Re}\tr(X^\dag Y)$.
Fact~\ref{item:small-ball-volumes} gives
\begin{equation}\label{eq:volume-ratio}
     \frac{V_{\mathrm F}}{V_{\mathrm{op}}}
 =\frac{\prod_{i=1}^{\alpha}\prod_{j=1}^{d'}(i+j-1)}{n!}
 \leq\left(\frac{e(\alpha+d')}{2n}\right)^n.
\end{equation}
Here AM--GM bounds the numerator by
$((\alpha+d')/2)^n$, and $n!\geq(n/e)^n$.
By Facts~\ref{item:small-ball-distance} and~\ref{lem:haar-block-law},
$$
 \begin{aligned}
 \mu\{Q:\|P-Q\|_1\leq s\}
 &=\Pr\{\|B\|_1\leq t\}\\
 &=\frac{\operatorname{Vol}(\mathcal B_{\mathrm{op}}\cap t\mathcal B_*)}
          {V_{\mathrm{op}}}\\
 &\leq\min\left\{1,\,
       t^{2n}\frac{\operatorname{Vol}(\mathcal B_*)}{V_{\mathrm{op}}}
       \right\}\\
 &\leq\min\left\{1,\,
       t^{2n}\left(\frac{V_{\mathrm F}}{V_{\mathrm{op}}}\right)^2
       \right\}\\
 &\leq\min\left\{1,\,
       \left(\frac{e(\alpha+d')s}{4\alpha d'}\right)^{2\alpha d'}
       \right\}.
 \end{aligned}
$$
The penultimate inequality uses Fact~\ref{item:small-ball-polarity};
the last uses the volume-ratio estimate  derived in \cref{eq:volume-ratio} and $t=s/2$.
\end{proof}

\section{Query lower bounds}\label{sec:query-lower}
Here, we combine the hardness of the discrimination problem in \cref{thm:hard-isom-disc} with the existence of packing nets in \cref{thm:packing} to prove the lower bounds.

\begin{theorem}[Near-boundary query lower bounds]
\label{thm:near-boundary-query-lower}
Let $d_1,d_2\ge 2$ and $r\ge 1$ be integers satisfying
$d_1<rd_2<\frac{4}{3}d_1$.
Write $\tau=rd_2/d_1$, $\alpha=rd_2-d_1$, and
$\chi=\min\{\alpha d_1,\alpha^2d_2^2\}$.
Suppose an algorithm, given query access to any unknown
$\calE\in\qchannel_{d_1,d_2}^r$, makes $n$ queries and
outputs an estimate $\widehat{\calE}$.
\begin{itemize}
\item Assume $0<\varepsilon\leq 1/(8e^{3/2}\sqrt8)$. If $\|\widehat{\calE}-\calE\|_\diamond\le\varepsilon$ with probability at least $2/3$, then
\[
  n\geq \frac{1}{3072e^3}\frac{rd_1d_2}{\varepsilon^2}.
\]
\item Assume $0<\varepsilon\leq10^{-12}$. If $\frac1{d_1}\|C_{\widehat{\calE}}-C_{\calE}\|_1\le\varepsilon$ with probability at least $2/3$, then
\[
 n\geq\frac{1}{2\cdot10^{28}}\left(
    \frac{d_1^2}{\varepsilon}
    +\frac{
      \min\{(\tau-1)d_1^2,(\tau-1)^2d_1^2d_2^2\}
    }{\varepsilon^2}
  \right).
\]
\end{itemize}
\end{theorem}
\begin{proof}
%Put
%\[
%d'=\lfloor d_1/2\rfloor,\qquad
%m=\alpha d',\qquad b=d'(d'+\alpha).
%\]
By \cref{eq:dim-inequ},
\begin{equation}\label{eq:lower-bound-b}
\frac{rd_1d_2}{6}\le d'(d'+\alpha)<\frac{3rd_1d_2}{8},
\qquad\text{and hence}\qquad d'(d'+\alpha)=\Theta(d_1^2).
\end{equation}

We first give a common reduction from tomography to discrimination.
Fix $0<\eta<1$, and let $\calN$ be the packing given by
\cref{thm:packing}. For each $P\in\calN$, write
$\calO_P=\{\calE_{P,U}:U\in\mathbb U_{d'}\}$.
Suppose that an $n$-query tomography algorithm outputs an estimate
$\widehat{\calE}$ within distance $\varepsilon$ of the unknown
channel, with probability at least $2/3$, where the distance is either
  Choi trace norm or diamond norm. If the distance between
any two distinct families $\calO_P$ and $\calO_Q$ is strictly greater
than $2\varepsilon$, then the minimum-distance decoder
\[
\widehat P\in\argmin_{Q\in\calN}\ 
\inf_{\calF\in\calO_Q}
\operatorname{dist}(\widehat{\calE},\calF)
\]
identifies $P$ whenever tomography succeeds, by the triangle
inequality. Thus it identifies uniformly random $P\in\calN$, with an
independent Haar-random mask $U\in\mathbb U_{d'}$, with probability at
least $2/3$.

By \cref{prop:hard-family}, all these channels have Kraus rank $r$.
Moreover, each query to $\calE_{P,U}$ can be simulated by a query to its
Stinespring isometry $V_{P,U}$ followed by tracing out the ancilla.
The latter isometry family is the one in \cref{prob-10260459}, up to
an output space basis change. Therefore \cref{thm:hard-isom-disc}, together
with the packing cardinality in \cref{eq:packing-size}, gives
\begin{equation}\label{eq:lower-bound-success}
\frac23\le \Pr[\widehat P=P]
\le \frac1{|\calN|}\left(e+\frac{2en}{d'(d'+\alpha)}\right)^{\alpha d'}
\le 2\left(e\eta^2+\frac{2en\eta^2}{d'(d'+\alpha)}\right)^{\alpha d'}.
\end{equation}
In particular, if $\eta\le 1/(4\sqrt e)$ and
$n\le d'(d'+\alpha)/(16e\eta^2)$, the right-hand side of
\cref{eq:lower-bound-success} is at most
$2(3/16)^{\alpha d'}\le 3/8<2/3$, a contradiction. Consequently,
\begin{equation}\label{eq:lower-bound-general}
n>\frac{d'(d'+\alpha)}{16e\eta^2}
\end{equation}
whenever $\eta\le 1/(4\sqrt e)$ and the corresponding packing
separation exceeds $2\varepsilon$.

\emph{Diamond norm.}
Choose $\eta=2e\sqrt8\,\varepsilon$. By
\cref{eq:packing-diamond}, distinct families are separated by more
than $2\varepsilon$ in diamond norm. By the assumed bound on
$\varepsilon$, we have $\eta\le1/(4\sqrt e)$, so
\cref{eq:lower-bound-general,eq:lower-bound-b} yield
\[
n>\frac{d'(d'+\alpha)}{512e^3\varepsilon^2}
\geq\frac{rd_1d_2}{3072e^3\varepsilon^2},
\]
proving the near-boundary lower bound in \cref{tab:diamond-bounds}.

\emph{Choi trace norm.}
Choose
\[
\eta=\frac{6e\sqrt{C_0}\,d_1}{\sqrt \chi}\,\varepsilon,
\qquad C_0=450.
\]
By \cref{eq:packing-choi}, the corresponding  Choi trace
norm separation is strictly greater than $2\varepsilon$.
If
\begin{equation}\label{eq:lower-bound-small-eps}
\varepsilon\le\frac{\sqrt \chi}{24e^{3/2}\sqrt{C_0}\,d_1},
\end{equation}
then $\eta\le1/(4\sqrt e)$, and
\cref{eq:lower-bound-general,eq:lower-bound-b} imply
\begin{equation}\label{eq:lower-bound-choi-classical}
n>\frac{d'(d'+\alpha)\chi}{576e^3C_0d_1^2\varepsilon^2}
\geq\frac{\chi}{3456e^3C_0\varepsilon^2}.
\end{equation}

On the other hand, \cref{cor:uniform-heisenberg}  gives for every dimension in the present regime and for    Choi trace norm  tomography,
\begin{equation}\label{eq:lower-bound-choi-heisenberg}
n\geq 10^{-28}\frac{d_1^2}{\varepsilon}.
\end{equation}
If \cref{eq:lower-bound-small-eps} does not hold, then
$$
\frac \chi{\varepsilon^2}<576e^3C_0d_1^2
\leq\frac{d_1^2}{\varepsilon},
$$
where the last inequality uses $576e^3C_0\varepsilon\leq1$, which follows from $C_0=450$ and $\varepsilon\leq10^{-12}$. Thus \cref{eq:lower-bound-choi-heisenberg} also controls the classical term in this case. If \cref{eq:lower-bound-small-eps} holds, both lower bounds apply and $1/(3456e^3C_0)>10^{-28}$.
Therefore, we can always combine \cref{eq:lower-bound-choi-classical,eq:lower-bound-choi-heisenberg} to prove
\[
n\geq\frac{1}{2\cdot10^{28}}\left(\frac{d_1^2}{\varepsilon}
+\frac{\chi}{\varepsilon^2}\right)
=\frac{1}{2\cdot10^{28}}\left(\frac{d_1^2}{\varepsilon}
+\frac{\min\{(\tau-1)d_1^2,(\tau-1)^2d_1^2d_2^2\}}
{\varepsilon^2}\right).
\]
\end{proof}

\section*{AI Disclosure}
\paragraph{Choi trace norm.}
The authors first considered isometry tomography and established a lower bound $\Omega(d^2/\varepsilon+(D-d)d/\varepsilon^2)$ under the diamond norm by providing GPT-6 with a draft introducing an isometry discrimination task involving a Haar-random mask, which generalizes the framework of \cite{chen2026quantum}. After multiple rounds of interaction with GPT-6, the authors extended this lower bound to the   Choi trace norm by refining the discrimination instances.
The authors then directed GPT-6 to extend the analysis to general channels. After several iterations, GPT-6 established a lower bound $\Omega(d_1^2/\varepsilon+\min\{(\tau-1)d_1^2,(\tau-1)^2d_1^2d_2^2\}/\varepsilon^2)$ and also demonstrated its optimality by proving a matching upper bound.

\paragraph{Diamond norm.}
The authors also first considered isometry tomography and directed GPT-6 to extend the parallel-query upper bound for diamond norm unitary tomography~\cite{he2026query} to isometry tomography, using the results on Choi norm isometry tomography~\cite{yoshida2025quantum} as guidance.
GPT-6 showed that the proposed generalization fails and instead established a lower bound $\Omega(Dd/\varepsilon^2)$, matching the existing upper bound~\cite{mele2025optimallearningquantumchannels,chen2026quantum}.
Then, by directing GPT-6 with the lower bound framework of \cite{chen2026quantum} (i.e., isometry discrimination with well-separated packing net), the lower bound is extended to $\Omega(rd_1d_2/\varepsilon^2)$ for general channels in the near-boundary regime.

\paragraph{Responsibility.}
The authors subsequently verified, learned, and revised all AI-generated arguments, and also reformulated and unified the lower bound proof for both norms.
The authors take full responsibility for the correctness and presentation of the results.

\bibliographystyle{alpha}
\bibliography{main}

\newcommand{\etalchar}[1]{$^{#1}$}
\begin{thebibliography}{SSKKG22}

\bibitem[BG73]{bjorck1973angles}
{\AA}ke Bj{\"o}rck and Gene~H. Golub.
\newblock Numerical methods for computing angles between linear subspaces.
\newblock {\em Mathematics of Computation}, 27(123):579--594, 1973.

\bibitem[BK15]{bianchi2015fourier}
Gabriele Bianchi and Michael Kelly.
\newblock A {Fourier} analytic proof of the {Blaschke--Santal\'o} inequality.
\newblock {\em Proceedings of the American Mathematical Society}, 143(11):4901--4912, 2015.

\bibitem[Bla17]{blaschke1917affine}
Wilhelm Blaschke.
\newblock {\"U}ber affine geometrie {VII}: Neue extremeigenschaften von ellipse und ellipsoid.
\newblock {\em Berichte {\"u}ber die Verhandlungen der K{\"o}niglich-S{\"a}chsischen Gesellschaft der Wissenschaften zu Leipzig, Mathematisch-Physische Klasse}, 69:306--319, 1917.

\bibitem[BLM13]{boucheron2013concentration}
St{\'e}phane Boucheron, G{\'a}bor Lugosi, and Pascal Massart.
\newblock {\em Concentration Inequalities: A Nonasymptotic Theory of Independence}.
\newblock Oxford University Press, 2013.

\bibitem[BM99]{bruss1999optimal}
Dagmar Bru{\ss} and Chiara Macchiavello.
\newblock Optimal state estimation for d-dimensional quantum systems.
\newblock {\em Physics Letters A}, 253(5-6):249--251, 1999.

\bibitem[BMQ21]{bavaresco2021strict}
Jessica Bavaresco, Mio Murao, and Marco~T{\'u}lio Quintino.
\newblock Strict hierarchy between parallel, sequential, and indefinite-causal-order strategies for channel discrimination.
\newblock {\em Physical Review Letters}, 127(20):200504, 2021.

\bibitem[BMQ22]{bavaresco2022unitary}
Jessica Bavaresco, Mio Murao, and Marco~T{\'u}lio Quintino.
\newblock Unitary channel discrimination beyond group structures: Advantages of sequential and indefinite-causal-order strategies.
\newblock {\em Journal of Mathematical Physics}, 63(4):042203, 2022.

\bibitem[BPGM26]{bravoprieto2026memory}
Carlos Bravo-Prieto, Weiyuan Gong, and Antonio~Anna Mele.
\newblock Quantum memory advantage for quantum process tomography.
\newblock {\em arXiv preprint arXiv:2607.13476}, 2026.

\bibitem[CDP08]{chiribella2008quantum}
Giulio Chiribella, G~Mauro D’Ariano, and Paolo Perinotti.
\newblock Quantum circuit architecture.
\newblock {\em Physical Review Letters}, 101(6):060401, 2008.

\bibitem[CDP09]{chiribella2009theoretical}
Giulio Chiribella, Giacomo~Mauro D’Ariano, and Paolo Perinotti.
\newblock Theoretical framework for quantum networks.
\newblock {\em Physical Review A}, 80(2):022339, 2009.

\bibitem[CGO{\etalchar{+}}26]{chen2026quantum}
Kean Chen, Filippo Girardi, Aadil Oufkir, Nengkun Yu, and Zhicheng Zhang.
\newblock Quantum channel tomography: optimal bounds and a {H}eisenberg-to-classical phase transition.
\newblock In {\em 67th IEEE Symposium on Foundations of Computer Science}, 2026.

\bibitem[Che25]{chen2025inverse}
Kean Chen.
\newblock Inverse-free quantum state estimation with {H}eisenberg scaling.
\newblock {\em arXiv preprint arXiv:2510.25750}, 2025.

\bibitem[Cho75]{choi1975completely}
Man-Duen Choi.
\newblock Completely positive linear maps on complex matrices.
\newblock {\em Linear Algebra and its Applications}, 10(3):285--290, 1975.

\bibitem[CN97]{chuang1997prescription}
Isaac~L Chuang and Michael~A Nielsen.
\newblock Prescription for experimental determination of the dynamics of a quantum black box.
\newblock {\em Journal of Modern Optics}, 44(11-12):2455--2467, 1997.

\bibitem[CO26]{chenoufkir2026incoherent}
Kean Chen and Aadil Oufkir.
\newblock Near-optimal incoherent tomography of low-rank quantum channels.
\newblock {\em arXiv preprint arXiv:2609.23752}, 2026.

\bibitem[CYZ25]{chen2025quantum}
Kean Chen, Nengkun Yu, and Zhicheng Zhang.
\newblock Quantum channel tomography and estimation by local test.
\newblock {\em arXiv preprint arXiv:2512.13614}, 2025.

\bibitem[CZY26]{chen2026optimal}
Kean Chen, Zhicheng Zhang, and Nengkun Yu.
\newblock Optimal lower bound for quantum channel tomography in away-from-boundary regime.
\newblock {\em arXiv preprint arXiv:2601.10683}, 2026.

\bibitem[DLP01]{d2001quantum}
G.~M. D'Ariano and P.~Lo~Presti.
\newblock Quantum tomography for measuring experimentally the matrix elements of an arbitrary quantum operation.
\newblock {\em Physical Review Letters}, 86(19):4195--4198, 2001.

\bibitem[EGH{\etalchar{+}}11]{etingof2011introduction}
Pavel Etingof, Oleg Golberg, Sebastian Hensel, Tiankai Liu, Alex Schwendner, Dmitry Vaintrob, and Elena Yudovina.
\newblock {\em Introduction to Representation Theory}, volume~59 of {\em Student Mathematical Library}.
\newblock American Mathematical Society, 2011.

\bibitem[FH13]{fulton2013representation}
William Fulton and Joe Harris.
\newblock {\em Representation Theory: A First Course}, volume 129 of {\em Graduate Texts in Mathematics}.
\newblock Springer, 2013.

\bibitem[FK07]{fyodorov2007colour}
Yan~V. Fyodorov and Boris~A. Khoruzhenko.
\newblock A few remarks on colour-flavour transformations, truncations of random unitary matrices, {Berezin} reproducing kernels and {Selberg}-type integrals.
\newblock {\em Journal of Physics A: Mathematical and Theoretical}, 40(4):669--700, 2007.

\bibitem[For06]{forrester2006conductance}
Peter~J. Forrester.
\newblock Quantum conductance problems and the {Jacobi} ensemble.
\newblock {\em Journal of Physics A: Mathematical and General}, 39(22):6861--6870, 2006.

\bibitem[GKKT20]{GKKT20}
Madalin Gu{\c{t}}{\u{a}}, Jonas Kahn, Richard Kueng, and Joel~A. Tropp.
\newblock Fast state tomography with optimal error bounds.
\newblock {\em Journal of Physics A: Mathematical and Theoretical}, 53(20):204001, 2020.

\bibitem[GMZ{\etalchar{+}}25]{Girardi2025Dec}
Filippo Girardi, Francesco~Anna Mele, Haimeng Zhao, Marco Fanizza, and Ludovico Lami.
\newblock Random {Stinespring} superchannel: converting channel queries into dilation isometry queries.
\newblock {\em arXiv preprint arXiv:2512.20599}, 2025.

\bibitem[Hal69]{halmos1969subspaces}
Paul~R. Halmos.
\newblock Two subspaces.
\newblock {\em Transactions of the American Mathematical Society}, 144:381--389, 1969.

\bibitem[Hay98]{hayashi1998asymptotic}
Masahito Hayashi.
\newblock Asymptotic estimation theory for a finite-dimensional pure state model.
\newblock {\em Journal of Physics A: Mathematical and General}, 31(20):4633--4655, 1998.

\bibitem[HHHH10]{hashimoto2010unitary}
T.~Hashimoto, A.~Hayashi, M.~Hayashi, and M.~Horibe.
\newblock Unitary-process discrimination with error margin.
\newblock {\em Physical Review A}, 81:062327, 2010.

\bibitem[HHJ{\etalchar{+}}17]{Haah_2017}
Jeongwan Haah, Aram~W. Harrow, Zhengfeng Ji, Xiaodi Wu, and Nengkun Yu.
\newblock Sample-optimal tomography of quantum states.
\newblock {\em IEEE Transactions on Information Theory}, 63(9):5628--5641, 2017.

\bibitem[HKOT23]{haah2023query}
Jeongwan Haah, Robin Kothari, Ryan O’Donnell, and Ewin Tang.
\newblock Query-optimal estimation of unitary channels in diamond distance.
\newblock In {\em 64th IEEE Symposium on Foundations of Computer Science}, pages 363--390, 2023.

\bibitem[HLS{\etalchar{+}}26]{he2026optimal}
Entong He, Zihao Li, Noam Scully, Sisi Zhou, and Yuxiang Yang.
\newblock Optimal classical shadow estimation of unitary channels at {H}eisenberg limit.
\newblock {\em arXiv preprint arXiv:2606.13638}, 2026.

\bibitem[HLY26]{he2026query}
Entong He, Zihao Li, and Yuxiang Yang.
\newblock Query-optimal unitary channel tomography in diamond distance with parallel access.
\newblock {\em arXiv preprint arXiv:2609.38145}, 2026.

\bibitem[Hua63]{hua1963harmonic}
Loo-Keng Hua.
\newblock {\em Harmonic Analysis of Functions of Several Complex Variables in the Classical Domains}, volume~6 of {\em Translations of Mathematical Monographs}.
\newblock American Mathematical Society, Providence, RI, 1963.
\newblock English translation of the original Chinese edition, Science Press, Peking, 1958.

\bibitem[Jam72]{jamiolkowski1972linear}
Andrzej Jamio{\l}kowski.
\newblock Linear transformations which preserve trace and positive semidefiniteness of operators.
\newblock {\em Reports on Mathematical Physics}, 3(4):275--278, 1972.

\bibitem[Leu00]{leung2000towards}
Debbie Wun~Chi Leung.
\newblock {\em Towards Robust Quantum Computation}.
\newblock PhD thesis, Stanford University, 2000.

\bibitem[MB26]{mele2025optimallearningquantumchannels}
Antonio~Anna Mele and Lennart Bittel.
\newblock Optimal learning of quantum channels in diamond distance.
\newblock {\em Nature Physics}, 2026.

\bibitem[Ner15]{neretin2015matrix}
Yurii~A. Neretin.
\newblock Matrix beta-integrals: An overview.
\newblock In Piotr Kielanowski, Pierre Bieliavsky, Anatol Odzijewicz, Martin Schlichenmaier, and Theodore Voronov, editors, {\em Geometric Methods in Physics}, Trends in Mathematics, pages 257--272. Birkh{\"a}user, Cham, 2015.

\bibitem[OG26]{oufkir2026improved}
Aadil Oufkir and Filippo Girardi.
\newblock Improved lower bounds for learning quantum channels in diamond distance.
\newblock {\em arXiv preprint arXiv:2601.04180}, 2026.

\bibitem[Ouf23]{Oufkir_2023}
Aadil Oufkir.
\newblock Sample-optimal quantum process tomography with non-adaptive incoherent measurements.
\newblock In {\em 2023 IEEE International Symposium on Information Theory}, pages 1919--1924, 2023.

\bibitem[OW16]{10.1145/2897518.2897544}
Ryan O'Donnell and John Wright.
\newblock Efficient quantum tomography.
\newblock In {\em 48th Annual ACM Symposium on Theory of Computing}, pages 899--912, 2016.

\bibitem[PCZ97]{poyatos1997complete}
J.~F. Poyatos, J.~I. Cirac, and P.~Zoller.
\newblock Complete characterization of a quantum process: the two-bit quantum gate.
\newblock {\em Physical Review Letters}, 78(2):390--393, 1997.

\bibitem[PSTW26]{pelecanos2025mixedstatetomographyreduces}
Angelos Pelecanos, Jack Spilecki, Ewin Tang, and John Wright.
\newblock Mixed state tomography reduces to pure state tomography.
\newblock In {\em 67th IEEE Symposium on Foundations of Computer Science}, 2026.

\bibitem[San49]{santalo1949invariante}
Luis~A. Santal{\'o}.
\newblock Un invariante af{\'i}n para los cuerpos convexos del espacio de $n$ dimensiones.
\newblock {\em Portugaliae Mathematica}, 8:155--161, 1949.

\bibitem[SSKKG22]{surawy2022projected}
Trystan Surawy-Stepney, Jonas Kahn, Richard Kueng, and Madalin Guta.
\newblock Projected least-squares quantum process tomography.
\newblock {\em Quantum}, 6:844, 2022.

\bibitem[SSW25]{scharnhorst2025optimal}
Thilo Scharnhorst, Jack Spilecki, and John Wright.
\newblock Optimal lower bounds for quantum state tomography.
\newblock {\em arXiv preprint arXiv:2510.07699}, 2025.

\bibitem[SZ26]{scully2026nearoptimal}
Noam Scully and Sisi Zhou.
\newblock Near-optimal parallel unitary process tomography in diamond distance.
\newblock {\em arXiv preprint arXiv:2609.36127}, 2026.

\bibitem[Ver18]{Ver18}
Roman Vershynin.
\newblock {\em High-Dimensional Probability: An Introduction with Applications in Data Science}.
\newblock Cambridge University Press, 2018.

\bibitem[Wat18]{watrous2018theory}
John Watrous.
\newblock {\em The Theory of Quantum Information}.
\newblock Cambridge University Press, 2018.

\bibitem[YMM26]{yoshida2025quantum}
Satoshi Yoshida, Jisho Miyazaki, and Mio Murao.
\newblock Quantum advantage in storage and retrieval of isometry channels.
\newblock {\em Physical Review Letters}, 136:190601, 2026.

\bibitem[YNU{\etalchar{+}}25]{yoshida2025random}
Satoshi Yoshida, Ryotaro Niwa, Takeru Utsumi, Ryuji Takagi, and Mio Murao.
\newblock Random dilation superchannel.
\newblock {\em arXiv preprint arXiv:2512.21260}, 2025.

\bibitem[YRC20]{yang2020optimal}
Yuxiang Yang, Renato Renner, and Giulio Chiribella.
\newblock Optimal universal programming of unitary gates.
\newblock {\em Physical Review Letters}, 125(21):210501, 2020.

\end{thebibliography}

\appendix
\section{A Heisenberg lower bound in all dimensions}\label{app:heisenberg}
In this section, we extend the Heisenberg lower bound in \cite[Theorem~6.1]{chen2026quantum}, which holds for sufficiently large dimensions $rd_1d_2\geq 7\cdot 10^{25}$, to arbitrary dimensions, including very small ones.

\begin{lemma}\label{lem:elementary-heisenberg}
Let $d_1,d_2\geq2$ and $r\geq1$ satisfy $rd_2\geq d_1$.
If an algorithm uses $n$ queries to learn every channel in
$\qchannel_{d_1,d_2}^r$ to Choi trace norm error
$0<\varepsilon\leq1/6$, with success probability at least $2/3$, then
\[
n\geq\frac{1}{9\varepsilon}.
\]
The same bound holds for diamond norm error.
\end{lemma}

\begin{proof}
Fix orthonormal bases $\{\ket{b}_{\mathrm B}\}_{b=1}^{d_2}$ of
$\mathcal H_{\mathrm B}$ and $\{\ket{e}_{\mathrm{anc}}\}_{e=1}^r$
of $\mathcal H_{\mathrm{anc}}$. For each $1\leq e\leq r$ and
$1\leq t\leq\lfloor d_2/2\rfloor$, consider the two vectors
\[
\ket{2t-1}_{\mathrm B}\otimes\ket{e}_{\mathrm{anc}},
\qquad
\ket{2t}_{\mathrm B}\otimes\ket{e}_{\mathrm{anc}}.
\]
These give $r\lfloor d_2/2\rfloor$ pairs of distinct orthonormal
vectors. Let
\[
L=\min\left\{
\left\lfloor\frac{d_1}{2}\right\rfloor,
r\left\lfloor\frac{d_2}{2}\right\rfloor
\right\}
\geq\frac{d_1}{3},
\]
where the inequality follows from $\lfloor k/2\rfloor\geq k/3$
for integers $k\geq2$ and $rd_2\geq d_1$.
Choose $L$ of these pairs and write them as
\[
\ket{a_j}=\ket{u_j}_{\mathrm B}\otimes\ket{e_j}_{\mathrm{anc}},
\qquad
\ket{b_j}=\ket{v_j}_{\mathrm B}\otimes\ket{e_j}_{\mathrm{anc}},
\qquad 1\leq j\leq L.
\]
Since $rd_2\geq d_1$, we can choose another $d_1-2L$ distinct
unused vectors $\ket{c_k}$ from the same product basis. Define
$V_0:\mathcal H_{\mathrm A}\to
\mathcal H_{\mathrm B}\otimes\mathcal H_{\mathrm{anc}}$ by
\[
\begin{aligned}
V_0\ket{2j-1}_{\mathrm A}&=\ket{a_j},\qquad
V_0\ket{2j}_{\mathrm A}=\ket{b_j}
&& (1\leq j\leq L),\\
V_0\ket{k}_{\mathrm A}&=\ket{c_k}
&& (2L<k\leq d_1).
\end{aligned}
\]
The images of the input basis vectors are orthonormal, so $V_0$
is an isometry. Set
\[
P_0=\sum_{j=1}^L\ket{2j}_{\mathrm A}\bra{2j}_{\mathrm A},
\qquad
U_\theta=e^{\mathrm i\theta P_0},
\qquad
V_\theta=V_0U_\theta,
\]
and define
\[
\calE_\theta(X)=\tr_{\mathrm{anc}}\bigl(
V_\theta X V_\theta^\dagger\bigr).
\]
Each $\calE_\theta$ belongs to $\qchannel_{d_1,d_2}^r$.

For each $j$, consider the following two basis vectors
of $\mathcal H_{\mathrm B}\otimes\mathcal H_{\mathrm A}$:
\[
\ket{u_j}_{\mathrm B}\otimes\ket{2j-1}_{\mathrm A},
\qquad
\ket{v_j}_{\mathrm B}\otimes\ket{2j}_{\mathrm A}.
\]
The $2\times2$ principal block of $C_{\calE_\theta}$ indexed by
these two vectors is
\[
\begin{pmatrix}
1&e^{-\mathrm i\theta}\\
e^{\mathrm i\theta}&1
\end{pmatrix}.
\]
The corresponding block of $C_{\calE_\theta}-C_{\calE_0}$ is therefore
\[
\begin{pmatrix}
0&e^{-\mathrm i\theta}-1\\
e^{\mathrm i\theta}-1&0
\end{pmatrix},
\]
which has eigenvalues $\pm2\sin(\theta/2)$ for
$0\leq\theta\leq1$, and hence trace norm $4\sin(\theta/2)$.
The $L$ pairs of basis vectors span mutually orthogonal subspaces.
For a Hermitian matrix, the sum of the trace norms of its diagonal
blocks on orthogonal subspaces does not exceed the trace norm
of the full matrix. Consequently,
\begin{equation}\label{eq:elementary-choi-gap}
\frac{1}{d_1}\|C_{\calE_\theta}-C_{\calE_0}\|_1
\geq\frac{4L}{d_1}\sin(\theta/2)
\geq\frac43\sin(\theta/2).
\end{equation}

A global phase on $V_\theta$ does not change $\calE_\theta$.
Thus, for $0\leq\theta\leq1$,
\begin{equation}\label{eq:elementary-diamond-gap}
\begin{aligned}
\|\calE_\theta-\calE_0\|_\diamond
&\leq2\|e^{-\mathrm i\theta/2}V_\theta-V_0\|_{\mathrm{op}}\\
&=2\|e^{-\mathrm i\theta/2}U_\theta-I_{\mathrm A}\|_{\mathrm{op}}\\
&=4\sin(\theta/4)\leq\theta.
\end{aligned}
\end{equation}

Take $\theta=6\varepsilon\leq1$. Since
$\sin x\geq2x/\pi$ for $0\leq x\leq\pi/2$,
\cref{eq:elementary-choi-gap} gives
\[
\frac{1}{d_1}\|C_{\calE_\theta}-C_{\calE_0}\|_1
\geq\frac{8\varepsilon}{\pi}>2\varepsilon.
\]
Therefore, tomography algorithm to within   Choi trace norm error $\varepsilon$ can distinguish $\mathcal{E}_0$ and $\mathcal{E}_\theta$ with success
probability at least $2/3$.

Let $\rho_0$ and $\rho_\theta$ be the algorithm's final states
under $\calE_0$ and $\calE_\theta$, respectively.
Replacing the $n$ queries one at a time and using trace-norm
contractivity of the intervening operations gives
\[
\|\rho_\theta-\rho_0\|_1
\leq n\|\calE_\theta-\calE_0\|_\diamond
\leq n\theta.
\]
By the binary discrimination bound,
\[
\frac23\leq\frac12+\frac14\|\rho_\theta-\rho_0\|_1
\leq\frac12+\frac{n\theta}{4},
\]
so $n\geq2/(3\theta)=1/(9\varepsilon)$.
Finally, the Choi trace-norm distance is at most
diamond-norm distance, which gives the same lower bound for
diamond-norm tomography.
\end{proof}

\begin{corollary}[Heisenberg lower bound in all dimensions]\label{cor:uniform-heisenberg}
Let $d_1,d_2\geq2$, $r\geq1$, and $d_1\leq rd_2\leq4d_1/3$. For $0<\varepsilon\leq10^{-12}$, tomography of $\qchannel_{d_1,d_2}^r$ to   Choi trace norm or diamond norm error $\varepsilon$, with success probability at least $2/3$, requires
$$n\geq10^{-28}\frac{d_1^2}{\varepsilon}.$$
\end{corollary}
\begin{proof}
If $rd_1d_2\geq7\cdot10^{25}$, apply \cite[Theorem~6.1]{chen2026quantum} with its error parameter set to $3966000\varepsilon$. Its precision assumption holds because $3966000\varepsilon\leq1/39660$. The resulting lower bound is
$$
n\geq\frac{d_1^2}{(2\cdot10^{19})\,3966000\varepsilon}
\geq10^{-28}\frac{d_1^2}{\varepsilon}.
$$
Otherwise, $d_1^2\leq rd_1d_2<7\cdot10^{25}$, and \cref{lem:elementary-heisenberg} gives
$$
n\geq\frac{1}{18\varepsilon}
\geq\frac{d_1^2}{126\cdot10^{25}\varepsilon}
\geq10^{-28}\frac{d_1^2}{\varepsilon}.
$$
\end{proof}
\end{document}